\documentclass
{bb}
\newcommand{\sty}{\scriptstyle}
\newcommand{\ssty}{\scriptscriptstyle}
\newcommand{\tsty}{\textstyle}
\newcommand{\be}{\begin{equation}}
\newcommand{\ee}{\end{equation}}
\newcommand{\lb}[1]{\label{#1}}

\newcommand{\obs}[1]{[#1]_{{\tsty {\ssty \mathrm{obs}}}}}
\newcommand{\bigobs}[1]{\left[#1\right]_{{\tsty {\ssty \mathrm{obs}}}}}
\newcommand{\gen}[1]{#1_{i}}

\newcommand{\dl}{d_{\ssty L}}
\newcommand{\da}{d_{\ssty A}}
\newcommand{\dg}{d_{\ssty G}}
\newcommand{\dz}{d_z}

\newcommand{\gama}{\gamma_{\ssty A}}

\newcommand{\Mbar}{\bar{M}}

\newcommand{\etal}{et al.\ }
\newcommand{\drm}{\mathrm{d}}
\newcommand{\df}{\mathrm{d}}
\usepackage{txfonts}
\usepackage{graphicx}
\usepackage{xcolor}

\begin{document}
\title{The Power Spectrum of Cosmological Number Densities}
\author{Amanda R.\ Lopes, \inst{1}\inst{2}\thanks{amandalopes1920@gmail.com}
        Marcelo B.\ Ribeiro~~\inst{3}\thanks{mbr@if.ufrj.br (Corresponding
	author)}
\and    William R.\ Stoeger~~\inst{4}\thanks{In memoriam (1943-2014)}}
\institute{Valongo Observatory, Universidade Federal do Rio de Janeiro, Rio
	de Janeiro, Brazil
\and
Department of Astronomy, Observat{\'o}rio Nacional, Rio de Janeiro, Brazil
%(present address)
\and
Physics Institute, Universidade Federal do Rio de Janeiro, Rio de
Janeiro, Brazil
\and
Vatican Observatory Research Group, Steward Observatory, University of
Arizona, Tucson, USA}
\date{}
\abstract{}
{This paper studies the cosmological power spectrum (PS) of the
differential and integral galaxy volume number densities, respectively
$\gamma_i$ and $\gamma_i^{*}$, constructed with the cosmological
distances $d_i$ $(i={\sty A},{\sty G},{\sty L},{\sty Z})$, where
$\da$ is the angular diameter distance, $\dg$ is the galaxy area
distance, $\dl$ is the luminosity distance and $\dz$ is the redshift
distance. Theoretical and observational quantities were obtained
in the Friedmann-Lema\^{\i}tre-Robertson-Walker (FLRW) spacetime
with a non-vanishing cosmological constant. The radial correlation
$\Xi_i$, a quantity defined in the context of these densities, is also
discussed in the wave number domain. All observational quantities were
computed using luminosity function (LF) data obtained from the FORS
Deep Field (FDF) galaxy survey.}
{The theoretical and observational power spectra of $\gamma_i$,
$\gamma_i^{\ast}$, $\Xi_i$ and the ratio $\gamma_i / \gamma_i^\ast$
were calculated by performing Fourier transforms on values of these
densities which were previously derived by Iribarrem \etal (2012)
from the observational values $\obs{\gamma}$ and $\obs{\gamma^\ast}$
obtained by using the galactic absolute magnitudes and Schechter's
parameters of the galaxy LF presented in Gabasch \etal (2004, 2006).
These parameters were evaluated from a I-band selected dataset of
the FDF in the redshift range $0.5 \le z \le5.0$ for its blue bands
and $0.75 \le z \le 3.0$ for its red ones.}
{The results show similar behavior of the power spectra obtained from
$\gamma$ and $\gamma^{\ast}$ using $\dl$, $\dz$ and $\dg$ as distance
measures. The PS of the densities defined with $\da$ have a different
and inconclusive behavior because this cosmological distance reaches a
maximum at $z\approx 1.6$ in the adopted cosmological model. For the
other distances our results suggest that the PS of $\obs{\gamma_i}$,
$\obs{\gamma^\ast_i}$ and $\obs{\gamma_i / \gamma^{\ast}_i}$
have a general behavior approximately similar to the power spectra
obtained with the galaxy two-point correlation function and, by being
sample size independent, they may be considered as alternative
analytical tools to study the galaxy distribution.}
{}
\titlerunning{The Power Spectrum of Cosmological Number Densities}
\authorrunning{Lopes et al.\ }
\maketitle

\section{Introduction}

The \textit{galaxy volume number densities} are important measures
in cosmology as they give information about the evolving universe
galaxy content and allow us to test cosmological models. This is so
because they are able to connect theory and observation once different
types of number densities are evaluated, both theoretically and
observationally. Theoretically, these densities can be obtained once
one chooses a cosmological spacetime geometry, usually the standard
model given by the spatially homogeneous
Friedmann-Lema\^{\i}tre-Robertson-Walker (FLRW) cosmology or, in
a recent trend, the spatially inhomogeneous Lema\^{\i}tre-Tolman-Bondi
cosmology as well (Iribarrem et al.\ 2013, 2014). Observationally,
inasmuch as the galaxy \textit{luminosity function} (LF) is a number
density per unit luminosity, several authors have been systematically
determining galaxy volume number densities using the LF derived from
galaxy redshift surveys.

Comparing cosmological models with observations using number
densities determined from LF parameters derived from observed data
requires, however, some model connecting relativistic cosmology
number counts theory to the LF astronomical data and practice. One
way of pursuing this theoretical link was advanced by Ribeiro \&
Stoeger (2003, hereafter RS03) who used general relativistic theory
to obtain specialized expressions capable of being applied to LF
data. Albani et al.\ (2007, hereafter A07) and Iribarrem et al.\
(2012, hereafter Ir12) further extended such a link and applied
the connecting expressions to the data of the Second Canadian
Network for Observational Cosmology (CNOC2; see Lin \etal 1999)
and FORS Deep Field (FDF; see Gabasch \etal 2004, 2006; hereafter
G04 and G06 respectively) galaxy redshift surveys. Following the
methodology advanced by Ribeiro (2005; see also Rangel Lemos \&
Ribeiro 2008) for number density definitions, where the adoption
of cosmological distance measures is made explicit, A07 and Ir12
calculated the \textit{differential number density} $\gamma$ and
its \textit{integral} counterpart $\gamma^{*}$ for various
cosmological distances and used them to study the observational
inhomogeneities in the relativistic radial distribution of
galaxies (Ribeiro 2001) belonging to both datasets in the FLRW
universe model. Their results indicated that observational
inhomogeneities in the relativistic galaxy distribution can arise
due to geometrical, past-light cone, effects even in the context
of spatially homogeneous cosmologies.

Analyzes of the large-scale structure of the universe use, however,
a complementary tool to examine such structures, the \textit{power
spectrum} (PS), which is basically a Fourier transform of the
quantity under study. If galaxies form the basic units of a cosmological
fluid described by a volume number density, this galactic fluid can
also be considered as being composed of wave densities with different
wave numbers. Then, the PS will give us the intensity, or power, in
each component. This is essentially the PS analysis of the galaxy
2-point correlation function, as the information given by the PS is
complementary to the correlation function. Various authors performed
such an analysis (e.g. Sylos Labini \etal 1998; Mart\'inez \& Saar
2002; Tegmark \etal 2004, 2006; Gabrielli \etal 2005) and showed
that, in general, the PS behavior is a power-law. Since galaxy
volume number densities were used by RS03, A07 and Ir12 to analyze
the universe galaxy content and evolution, it is natural to expand
these studies to include a PS analysis.

The goal of this work is to study the differential and integral
number densities $\gamma$ and $\gamma^\ast$ in the wave number domain
through the PS. This is achieved by performing Fourier transforms of
both quantities. Using the relativistic analysis and results presented
by Ir12, we computed the theoretical power spectra of $\gamma_i$ and
$\gamma_i^\ast$ obtained using the cosmological distance $d_i$. Here
the index $i$ stands for the adopted distance measure, namely the
\textit{angular diameter distance} $\da$, also known as \textit{area
distance}, \textit{galaxy area distance} $\dg$, \textit{luminosity
distance} $\dl$ and \textit{redshift distance}
$\dz$ $(i = {\sty A},{\sty G},{\sty L}\, ,{\sty Z})$. Observational
values of the differential and integral densities computed with these
distance measures, $\obs{\gamma_i}$ and $\obs{\gamma_i^\ast}$, were
obtained using the results of Ir12, which employed the LF parameters
derived by G04 and G06 from the FDF survey data. The analysis is also
extended to include the power spectra of the \textit{ratio}
$\gamma / \gamma^{*}$ and of another quantity, named here as
\textit{radial correlation} $\Xi$, which is basically a radial function
whose behavior in terms of $d_i$ bears some similarities to the
2-point correlation function (Ribeiro 1995).

The distance measures $d_i(z)$ were obtained in the cosmology
adopted here: the FLRW model with non zero cosmological constant
$\Lambda$ and parameter values equal to $\Omega_{m_{\ssty 0}}=0.3$, 
$\Omega_{\Lambda_{\ssty 0}} = 0.7$ and $H_0=70 \; \mbox{km} \;
{\mbox{s}}^{-1} \; {\mbox{Mpc}}^{-1}$. The dataset studied here is 
the I-band selected luminosity functions of the FDF survey. Ir12
calculated observed differential number counts and then 
$\obs{\gamma_i(z)}$ and $\obs{\gamma_i^\ast (z)}$ using the
absolute magnitudes, LF Schechter parameters and their respective
redshift parametrization, all previously derived by G04 and G06
directly from the FDF data in eight bandwidths with a redshift range
$0.5\le z \le 5.0$, for the blue bands (G04), and $0.45 \le z \le
3.75$ for the red bands (G06). Those results allowed us to
calculate the ratio $\obs{\gamma_i / \gamma_i^\ast}$ and the radial
correlation $\obs{\Xi_i(d_i)}$. Fourier transforms on these
quantities provided their respective power spectra,
$P_k \obs{\gamma_i}$, $P_k \obs{\gamma_i^\ast}$, $P_k \obs{\gamma_i
/ \gamma_i^\ast}$ and $P_k \obs{\Xi_i}$. 

The results show that power spectra analyzes with the angular diameter
distance $\da$ is problematic, because each $\da$ corresponds to two
different values for $z$, since this distance starts from zero, grows
to maximum and then tends to zero as $z \rightarrow \infty$. This 
maximum $\da$ value is the only one with unique redshift. Due to these
features the results obtained with $\da$ turn out to be problematic
in terms of interpretation. 

We were also able to reproduce the preliminary results of Ribeiro (1995)
regarding the radial correlation, confirming its dependency with the
sample size and finding out that its respective PS is also sample size
dependent. The best results, however, were obtained with the power
spectra of the differential and integral densities using $\dl$ and
$\dz$, which show a behavior bearing similarities, even at wave number
values, with the 2-point correlation function PS. This suggests that
PS analysis of number densities can also be a complementary tool for
large-scale structure studies.

The plan of the paper is as follows. In \S \ref{basic} the basic 
quantities used in this paper are presented, as well as the expressions
required in their calculation. \S \ref{flrw} specializes the previously
discussed general quantities to the FLRW cosmological model and shows
how the luminosity function parameters of the FDF survey are used to
calculated the desired observational expressions. \S \ref{results}
presents both the theoretical and observation results of all quantities
discussed here, as well as their respective power spectra, and \S
\ref{conclusions} presents our conclusions. 

\section{Basic Definitions}\label{basic}

This section defines the basic quantities and expressions necessary
for the PS analysis discussed here. Although they have already been
introduced in previous papers (Ribeiro 1995, 2005; Rangel Lemos \&
Ribeiro 2008; A07; Ir12), one can find below a few additional
clarifying remarks.

\subsection{Differential and Integral Densities}\lb{densi}

Let us start by defining the two key quantities used in this study.
The {\it differential density} $\gen{\gamma}$ gives the rate
of growth in number counts, or more precisely in their density, as one
moves along the observational distance $\gen{d}$. It is defined by
the following expression (Ribeiro 2005),
\begin{equation}
\label{gammadef}
\gen{\gamma}=\frac{1}{4 \pi (\gen{d})^2}\frac{\drm N}{\drm (\gen{d})},
\end{equation}
where $N$ is the \textit{cumulative number counts}. The {\it
integrated differential density}, or simply \textit{integral
density}, $\gen{\gamma}^\ast$ gives the number of sources per
unit of observational volume located inside the observer's past
light cone down to a distance $\gen{d}$. It is written as follows,
\be
\gen{\gamma^\ast}=\frac{1}{\gen{V}} \int\limits_{\gen{V}} \gen{\gamma}
\; \drm \gen{V},
\lb{gest}
\ee
where $\gen{V}$ is the {\it observational volume}, 
\begin{equation}
\lb{volume}
\gen{V}= \frac{4}{3} \pi (\gen{d})^3.
\end{equation}
As it has been previously discussed, these quantities are useful in
determining whether or not, and at what ranges, a spatially
homogeneous cosmological model exhibits observational inhomogeneity
because these densities behave very differently depending on the
distance measure used in their definitions. In other words,they
show a clear dependence on the adopted cosmological distance
(Ribeiro 2005; Rangel Lemos \& Ribeiro 2008; A07; Ir12).

It is useful to write the two number densities above in terms of the
redshift. Thus, the differential density (\ref{gammadef}) may be
written as,  
\begin{equation}
\label{gamma}
\gen{\gamma}=\frac{\drm N}{\drm z}\left\{ 4 \pi (\gen{d})^2 \;
\frac{\drm (\gen{d})}{\drm z} \right\} ^{-1},
\end{equation}
where $\drm N /\drm z$ is the \textit{differential number counts}.
In addition, the expressions above allow us to conclude that
the following expression holds,
\be
   \gen{\gamma^\ast} = \frac{N}{\gen{V}}.
   \lb{gest3}
\ee

\subsection{Radial Correlation}

Studies of the large-scale galaxy structure have traditionally
employed the 2-point correlation function as a tool to characterize
the galaxy distribution, including the possible depth at which this
distribution reached homogeneity. This latter application was,
however, criticized by Pietronero (1987) who argued that the
correlation function actually presupposes homogeneity and, therefore,
it is a tool unsuited to find its possible presence in the galaxy
structure. In order to remove such a built-in hypothesis, Pietronero
(1987) proposed a different correlation methodology by advancing the
conditional density $\Gamma(d)$, where $d$ is the distance, as the
proper tool capable of detecting the possible homogeneity in the
galaxy distribution. He also showed that $\Gamma$ has a simple
functional relationship to the 2-point correlation function $\xi$,
given by the following expression (see also Sylos Labini \etal 1998;
Gabrielli \etal 2005),
\begin{equation}
\xi(d) = \frac{\Gamma(d)}{\bar{n}(R)} - 1,
\label{corr-func}
\end{equation}
where $\bar{n}(R)$ is the mean density of the sample whose radius
$R$ defines the sample volume. This equation shows explicitly the
correlation function dependence to the sample depth given by $R$.

Ribeiro (1995) followed these ideas and defined a radial quantity
exhibiting the same sample dependence property, dubbed here as
\textit{radial correlation} $\Xi$. In the present context this
quantity may be written as follows,
\begin{equation}
\Xi_{i}(d_i) = \frac{\gamma_{i}(d_i)}{\gamma^{*}_{i}(R_{i})} - 1.
\label{radial-corr}
\end{equation}
As the expressions above shows, $\Xi$ is not a proper statistical
correlation, but simply a dimensionless radial function that can be
used in the context of number density distributions originated from
relativistic cosmology theory, and whose behavior is also dependent
on the sample depth, defined here at the radial distance $d_i=R_i$.
Ribeiro (1995) studied this quantity in the context of the
Einstein-de Sitter model using its theoretical distribution obtained
in terms of the luminosity distance and redshift. Here, however,
we shall employ the FLRW cosmology as well as the other distance
measures discussed above in order to derive observational values for
$\Xi_i$ using the FDF galaxy survey.

\subsection{Power Spectrum}\lb{PS-section}

If one assumes spherical symmetry, the PS of the 2-point correlation
function is given by, 
\begin{equation}
P_{k}(\xi) = 4\pi \int_{0}^{\infty} \frac{\sin(k x)}{k x}\ \xi(x)
\, x^{2} \, \drm x.
\label{PS-simetric}
\end{equation}
This function describes clustering in terms of the wave numbers $k$
and separates the clustering effects in different scales. To study
the differential densities in the wave number domain, we shall employ
a similar definition as above, which may be written as follows,
\begin{equation}
P_{k}(\gamma_{i}) = \int_{0}^{\infty} \frac{\sin(kx)}{kx}
   \gamma_{i}(x)\ x^{2} \ \drm  x.
\label{PS-gama-a}
\end{equation}
For astronomical observations contained in galaxy survey samples
one needs, however, to consider the finiteness of the sample size
and divide it in redshift bins so that the function above can
actually be plotted from real data. This means rewriting the
equation above as follows, 
\begin{equation}
P_{k}\obs{\gamma_{i}} = \int_{0}^{d_{i}(z_j)} \frac{\sin(k_{j}\
d_{i})}{k_{j}\ d_{i}} \gamma_{i}\ {d_{i}}^{2} \ \drm (d_{i}).
\label{PS-gama}
\end{equation}
Here $d_i(z)$ is the distance-redshift relation given by the adopted
cosmological model in a certain distance measure,
\be
k_{j} \equiv \frac{ 2\pi}{ d_{i}(z_{j})}, \; \; \;
\mbox{for} \; \; \; (j=1,\ldots,m),
\lb{kj}
\ee
where $m$ is the total number of redshift bins dividing the sample and
$z_j$ is the central redshift value of $j$-th bin. Replacing
$\gamma_{i}$ with its definition given in Eq.\ (\ref{gammadef}) leads
to the following expression,
\begin{equation}
P_{k}\obs{\gamma_{i}} = \frac{1}{4 \pi} \int_{0}^{d_{i}(z_j)}
\frac{\sin(k_{j} \ d_{i})}{k_{j}\ d_{i}} \frac{\drm N}{\drm (d_{i})}
\ \drm (d_{i}).
\label{PS-gama2}
\end{equation}
Similarly, The PS of the integral densities yields,
\begin{equation}
P_{k}\obs{\gamma^\ast_{i}} = \int_{0}^{d_{i}(z_j)} \frac{\sin (k_{j}
 \ d_{i})} {k_{j}\ d_{i}} \ \gamma^\ast_{i} \ {d_{i}}^{2} \ \drm (d_{i}),
\label{def-PS-n}
\end{equation}
or, using Eqs.\ (\ref{volume}) and (\ref{gest3}), it can also be
written as,
\begin{equation}
P_{k}\obs{\gamma^\ast_{i}} = \frac{3}{4 \pi} \int_{0}^{d_{i}(z_j)}
\frac{\sin(k_{j} \ d_{i})} {k_{j}\ {d_{i}}^{2}} \ N \ \drm (d_{i}).
\label{def-PS-n2}
\end{equation}

The angular diameter distance $\da$ is the quantity required in theory
to obtain the theoretical number counts according to the relativistic
cosmology theory developed by Ellis (1971). However, observationally
measuring cosmological distances depends on circumstances, that is, they
depend on the available observations. If one is only able to measure the
redshift and intrinsic luminosity then the luminosity distance is the
distance measure of choice to calculate from observations. On the other
hand, if one has observed intrinsic areas and redshift one is able to
derive only the angular diameter distance (see also Ribeiro 2005). Hence,
using different distance measures and deriving their respective power
spectra opens the way of applying the theory developed here to a wider
range of observational possibilities.

\section{Cosmologic and Sample Specific Expressions}\label{flrw}

The quantities defined above are general, that is, valid for any
cosmological model. In this section we shall adopt the FLRW cosmology
and calculate them in this specific universe model, as well as using
galaxy redshift survey observations.

As a simple examination of the equations above shows, we only require
cosmologic specific expressions for $d_i(z)$, $\drm (d_i) / \drm z$,
$N(z)$, $\drm N / \drm z$, $\obs{N}$ and $\obs{\drm N(z) / \drm z}$.
The first four quantities are theoretical and come from the chosen
spacetime geometry, whereas the last two ones come from a combination
of theoretical and observational results, as we shall see next.

\subsection{FLRW Cosmological Model}

In this cosmology we need to obtain the required quantities along
the past null cone, a task that is carried out numerically. As Ir12
discusses such a procedure in great details, what we shall present
next is just a summary of the results needed here.

The scale factor of the FLRW cosmology with non zero cosmological
constant $\Lambda$ can be written in terms of the radial coordinate
$r$ as follows, 
\begin{equation}
\label{SvsR}
\frac{\drm S}{\drm r} = - H_0 \left[ \frac{ (\Omega_{\Lambda_0})S^4
		- {S_0}^2(\Omega_0-1)S^2
                + (\Omega_{m_0}{S_0}^3)S
		}
		{c^2-{H_0}^2 {S_0}^2 (\Omega_0
		- 1)r^2} \right]^{\frac{1}{2}},
\end{equation}
where $\Omega_{m_{0}}$ and $\Omega_{\Lambda_0}$ are, respectively,
the matter and cosmological density parameters, $H_0$ is the Hubble
constant, $S_0$ is the current value of the scale factor, assumed 
to be equal to unity, and
\begin{equation}
\label{omega0}
\Omega_0 \equiv \Omega_{m_0} + \Omega_{\Lambda_0}. 
\end{equation}
To find solutions for $S(r)$ we shall use the following numerical
values for the parameters above: $\Omega_{m_{\ssty 0}} = 0.3$,
$\Omega_{\Lambda_{\ssty 0}} = 0.7$ and $H_0=70 \; \mbox{km} \;
{\mbox{s}}^{-1} \; {\mbox{Mpc}}^{-1}$. The  redshift $z$ can
be written as
\begin{equation}
\label{red}
1+z=\frac{S_0}{S},
\end{equation}
where it is clear that a numerical solution of the scale factor
also produces numerical solutions for $z(r)$. The  differential
number counts yields,
\begin{eqnarray}
\label{dNdz2}
\frac{\drm N}{\drm z} & = & \left( \frac{3 \; c \; \Omega_{m_0} H_0
 {S_0}^2}{2 G {\cal M}_g} \right) \times \nonumber \\
& & \times \left[ \frac{r^2 S^2}{ \sqrt{ 
		(\Omega_{\Lambda_0})S^4 - {S_0}^2
		(\Omega_0-1)S^2 + (\Omega_{m_0}{S_0}^3)S}} \right],
\end{eqnarray}
where $G$ is the gravitational constant, $c$ is the light speed and
${\cal{M}}_g$ is the average galactic rest mass, dark matter included.

As pointed out by A07 and Ir12, the details of the galaxy mass function
and how it evolves with the redshift are imprinted in the LF itself and
will be included in any observationally derived functions stemming
from the LF. For the theoretical quantities, we shall assume a constant
average galaxy rest mass with the working value of ${\cal M}_g \approx
10^{11} {M}_{\sun}$ based on the estimate by Sparke \& Gallagher (2000). 
To actually extract from the LF the implicit galaxy mass function so that
the function ${\cal M}_g(z)$ can be estimated and used in the theoretical
quantities is a problem already dealt with in Lopes \etal (2014).

\begin{figure*}
\begin{center}
\includegraphics[width=9.5cm]{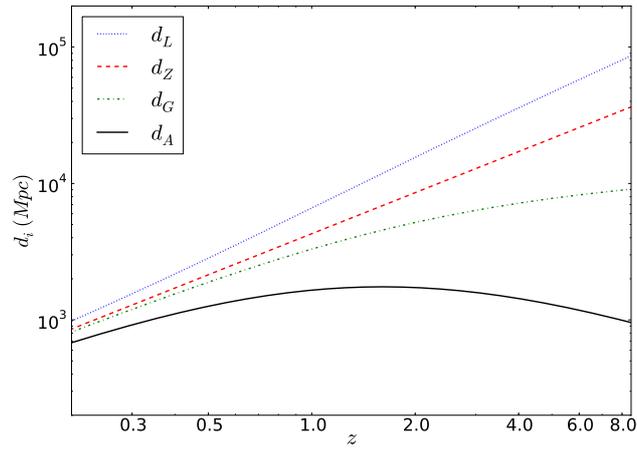}
\end{center}
\caption{Redshift evolution of the four FLRW cosmological distance
measures adopted here assuming $\Omega_{m}=0.3$, $\Omega_{\Lambda}=0.7$
and $H_{0}=70$ km s$^{-1}$ Mpc$^{-1}$. Symbols are as in the legend.}
\label{fig.distances}
\end{figure*}

The angular diameter distance $\da$ is obtained by means of a relation
between the intrinsically measured cross-sectional area element $\drm
\sigma$ of the source and the observed solid angle $\drm \Omega_0$.
In a spherically symmetric spacetime it may be written as follows
(Ellis 1971, 2007; Ribeiro 2005; Pleba\'{n}ski \& Krasi\'{n}ski 2006),
\begin{equation}
\label{da}
(\da)^2 = \frac{\drm \sigma}{\drm \Omega_0} =
(Sr)^2.
\end{equation}
The redshift distance is defined by the following expression,
\begin{equation}
\label{dz}
d_z = \frac{c \: z}{H_0},
\end{equation}
and the remaining distance measures can be obtained from the
area distance by means of the Etherington (1933) reciprocity
law (Ellis 1971, 2007), 
\begin{equation}
\label{rec}
\dl = (1+z)^2 \da = (1+z) \; \dg.
\end{equation}
The particular case of this law relating only $\dl$ to $\da$ is
known as the distance duality relation and can be determined
observationally (Holanda et al.\ 2010, 2011, 2012). Thus,
considering Eqs.\ (\ref{red}), (\ref{da}) and (\ref{rec}) the
other cosmological distances are straightforwardly written in
terms of the scale factor as follows,
\be \dl = {S_0}^2 \left( \frac{r}{S} \right), \lb{dl} \ee
\be \dg= S_0 \: r, \lb{dg} \ee
\be d_z=\frac{c}{H_0} \left( \frac{S_0}{S} -1 \right). \lb{dzS} \ee
Fig.\ \ref{fig.distances} shows the redshift evolution of these
distances in the adopted FLRW spacetime.

The derivatives of each distance measure in terms of the redshift
are also necessary. Remembering Eqs.\ (\ref{SvsR}), (\ref{red})
and (\ref{da}) one can easily conclude that, 
\begin{equation}
\label{diffda}
\frac{\drm (\da)}{\drm z} = \frac{\drm S}{\drm z} \frac{\drm r}{\drm S}
\frac{\drm  (\da)}{\drm r}= - \frac{S^2}{S_0} \left[r + S { \left(
\frac{\drm S}{\drm r} \right) }^{-1} \right].
\end{equation}
Similarly, both $\drm (\dl) / \drm z$ and $\drm (\dg) / \drm z$ are
easily obtained, whereas $\drm (\dz) / \drm z$ comes directly from
Eq.\ (\ref{dz}).

Finally, the cumulative number counts $N$ is derived as follows.
The number of cosmological sources per proper volume unit $n$ is
related to the matter density $\rho_m$ by means of, 
\begin{equation}
n = \frac{N}{V_{\ssty Pr}}= \frac{\rho_m}{{\cal M}_g},
\lb{n1}
\end{equation}
where $V_{\ssty Pr}$ is the proper volume. As shown by Ir12, in
the FLRW cosmology the right-hand side of the Einstein equations 
allows us to write the number density in terms of the proper volume
as follows,
\begin{equation}
\label{n}
n = \left( \frac{3 \Omega_{m_0} {H_0}^2 {S_0}^3}{8 \pi G {\cal M}_g}
    \right) \frac{1}{S^3}.
\end{equation}
In what follows we shall need to write the number density in terms of
the comoving volume $V_{\ssty C}$ and, therefore, a conversion between
$V_{\ssty Pr}$ and $V_{\ssty C}$ is needed. In the FLRW model this is
given by $V_{\ssty Pr} / V_{\ssty C} = S^3$, since $V_{\ssty C}=(4/3)
\pi r^3$ and $V_{\ssty Pr}=(4/3) \pi r^3 S^3$.

\subsection{Observational Quantities}\lb{obs}

The next step is to specialize the expressions above to obtain 
observational counterparts based on the LF data of a specific
galaxy catalog. To do so we need to use the methodology advanced
by RS03 connecting theoretical expressions to astronomically
derived quantities. 

In summary, assuming that an observational quantity $\obs{T}$
can be related to its theoretical counterpart $T$ by means of a
\textit{consistency function} $J$ such that $\obs{T} = J \, T$,
then the relationship between the {\em selection function}
$\psi$, which gives the volume number density of galaxies with
luminosity above a given threshold, to the comoving volume number
density $n_{\ssty C}\,(z)$ can be written as follows,
\be
\lb{Jdef2}
\psi (z) = J(z) \, n_{\ssty C}(z).
\ee
The redefinition of the number density in terms of comoving volume
is a consequence of the fact that in observational cosmology this
is the most adopted volume definition, rather than the proper volume
commonly found in theoretical calculations. Hence,
\begin{equation}
\label{radn}
n_{\ssty C} = \frac{N}{V_{\ssty C}} = \frac{3\, N}{4 \, \pi \, r^3}.
\end{equation}
What Eq.\ (\ref{Jdef2}) tells us is that the consistency
function basically represents the undetected mass fraction in relation
to the one predicted by theory. It also includes the galaxy redshift
survey data, since the {\em selection function} (SF) is written in terms of
the LF $\phi$ as follows,
\be
\lb{psi1}
\psi (z) = \int ^{\infty}_{l_{\mathrm{\ssty lim}}(z)} \phi(l) \, \drm l,
\ee
where $l_{\mathrm{\ssty lim}}(z)$ is the lower luminosity threshold
below which cosmological sources are not observed. 

The galaxy LF is a number density per unit of luminosity, obtained
by fitting the galaxy distribution in a given dataset to some
function and determining its parameter values. The most common
analytical form used to fit galaxy survey data is the Schechter
(1976) function, given as below, 
\begin{equation}
\lb{schechter}
\phi (l) = \phi^\ast \, l^{\, \alpha} \, e^{-l}.
\end{equation}
Here $l \equiv L/L_{\ssty *}$, $L$ is the observed luminosity,
$L_{\ssty *}$ is the luminosity scale parameter, $\phi^\ast$ is a
normalization parameter and $\alpha$ gives the faint-end slope
parameter. It was found by several works that the LF evolves with
the redshift (Iribarrem et al.\ 2012, 2013, 2014 and references
therein).

The observational quantities of interest studied in this paper
require previous knowledge of the observed differential number
counts $\obs{dN/dz}$. Therefore, linking this quantity to its 
theoretical counterpart is an essential step, yielding (Ir12), 
\begin{equation}
\label{Jdiffza}
\bigobs{\frac{\drm N}{\drm z}} = J(z) \frac{\drm N}{\drm z} =
 \frac{\psi(z)}{n_{\ssty C}(z)} \frac{\drm N}{\drm z}.
\end{equation}
This is our key equation relating the relativistic theory to the
observations. Thus, the observed cumulative number counts can be
calculated as,
\begin{equation}
\lb{obsN}
\obs{N(z)} = \int_0^z\bigobs{\frac{\drm N}{\drm z'}}\drm z',
\end{equation}
and, as discussed in Ir12, the uncertainty in $\obs{N}$ can be
estimated from the uncertainty in $\obs{\drm N/\drm z}$. 

Note that the approach summarized above indicates that all
observational quantities obtained by applying $J(z)$ to their
theoretical counterparts will inherit the same empirical number
count redshift evolution encoded in the parametrization of the
LF. In addition, Eq.\ (\ref{Jdef2}) shows that the consistency
function is independent of volume units and, hence, if an
observational quantity is obtained using $J(z)$ its original volume
unit dependence is preserved.  

One must also point out that the consistency function
$J(z)$ is characterized by the luminosity limits of the SF, but not only
that because besides representing its integration limit the SF also
represents the LF parameters themselves, which were obtained for a
specific dataset, so much so that they vary depending on the filters used
during observation and according to the specific galaxy types. The LF is
estimated for a given filter and for distinct types of galaxies in
specific environments, hence the SF inherits the same characteristics and
limitations of the LF. Therefore, when one estimates $J(z)$ from a given
SF its value will reflect not only the luminosity limit but also all other
observational limitations.

Finally, as discussed at length in RS03, the consistency
function reflects the differences between observational results and
theoretical predictions. Hence, its value should always be in the interval
$0\le J\le1$, where the unit means perfect agreement between theory and
observations. If $J>1$ that implies in a theory unable to explain the
observations and, therefore, the model must be revised. The case $J<1$ is
to be expected since our observational techniques are limited. From Eq.\
(\ref{Jdef2}) one can see that $J$ represents the undetected, observational
wise, fraction of objects in relation to what is predicted by theory. 

\subsection{FORS Deep Field Galaxy Redshift Survey}\label{fdf}

As seen above, the consistency function is required for the calculation
of the observational quantities, but to do so one needs first to compute
the SF in terms of the redshift in a given galaxy
redshift survey. That was done by Ir12, where one can find a very
thorough discussion of the procedure adopted in the calculation of
the FDF SFs, as well as details of the numerical
scheme. Hence, what we shall present below is a summary of Ir12's
methodology on this respect.

G04 and G06 fitted the LF Schechter parameters over the redshifts of
5558 I-band selected galaxies in the FORS Deep Field dataset,
photometrically measured down to an apparent magnitude limit of
$I_{\ssty AB}=26.8$. All galaxies were selected in the I-band and
then had their magnitudes for each of the five blue bands
(1500 \AA, 2800 \AA, $u'$, $g'$ and $B$) and the three red ones ($r'$,
$i'$ and {$z'$}) computed using the best fitting SED given by the 
photometric redshift code developed by the authors convolved with the
associated filter function. G04 and G06 determined the photometric
redshifts by fitting template spectra to the measured fluxes on the
optical and near infrared images of the galaxies. The redshift ranges
of $0.75 \le z \le 3.0$ for the red bands and $0.5 \le z \le 5.0$ for
the blue bands. 

To calculate the actual values of the SFs at each 
waveband, we need the result obtained in RS03 for the limited
bandwidth version of the SF of a given LF, fitted
by a Schechter's analytical profile (Eqs.\ \ref{psi1} and
\ref{schechter}). In terms of the absolute magnitudes $M(z)$
this expression reads,
\begin{eqnarray}
\lb{sf}
\psi^{\ssty W}(z) = 0.4 \ln 10 \, \int_{-\infty}^{M^{\ssty
W}_{\mathrm{\ssty lim}}(z)}
\phi^\ast(z) 10^{0.4[1+\alpha(z)][M^\ast(z) - \Mbar^{\ssty W}]}
\times \nonumber \\ \times
\exp \{-10^{0.4[M^\ast(z) - \Mbar_{\ssty W}]}\} \,
\drm \Mbar^{\ssty W},
\end{eqnarray}
where the index $W$ indicates the bandwidth filter. The expressions
for the redshift evolution of the LF parameters adopted by G04 and
G06 are,
\begin{eqnarray*}
\phi^{\ast}(z) = \phi^{\ast}_{\ssty 0} \, (1+z)^{B^{\ssty W}}, \\
M^{\ast}(z) = M^{\ast}_0 + A^{\ssty W} \ln (1+z), \\
\alpha(z) = \alpha_{\ssty 0},
\end{eqnarray*}
where $A^{\ssty W}$ and $B^{\ssty W}$ are the evolution parameters
fitted for the different $W$ bands and $M^{\ast}_{\ssty 0}$,
$\phi^{\ast}_{\ssty 0}$ and $\alpha_{\ssty 0}$ the local (z $\approx$ 0)
values of the Schechter parameters. Since all galaxies were detected
and selected in the I-band, we can write,
\begin{equation}
\label{mlim}
M^{\ssty W}_{\mathrm{\ssty lim}}(z) = M^{\ssty I}_{\mathrm{\ssty
lim}}(z) = I_{\mathrm{\ssty lim}} - 5 \log[\dl (z)] - 25 + A^{\ssty I},
\end{equation}
for a luminosity distance $\dl$ given in Mpc. $I_{\mathrm{\ssty
lim}}=26.8$ is the limiting apparent magnitude of the I-band in the
FDF survey. Its reddening correction is $A^{\ssty I} = 0.035$ (Heidt
\etal 2001).

The SFs for all eight bands of the dataset were
calculated by Ir12 using simple numerical integrations at equally
spaced values spanning the whole redshift interval using parameter
values as given by G04 and G06. The UV bands, 1500 \AA, 2800\AA, and
$u'$, evolve tightly with redshift, having values that are consistent
with each other within the uncertainties, while assuming values outside
the uncertainties of the ones calculated in the blue optical bands
$g'$, and $B$. Due to this, Ir12 chose to use the SFs of the combined
UV bands and those of the combined blue optical bands separately. The
SFs in the red-band dataset of G06, $r'$, $i'$ and $z'$ were also
combined. Once the combined SFs were obtained, Ir12 calculated
$\obs{\drm N/\drm z}$ by means of Eq.\ (\ref{Jdiffza}). 

\section{Power Spectrum Analysis of the FDF Survey}\label{results}

\begin{figure*}
\begin{center}$
\begin{array}{cc}
\includegraphics[width=9.5cm]{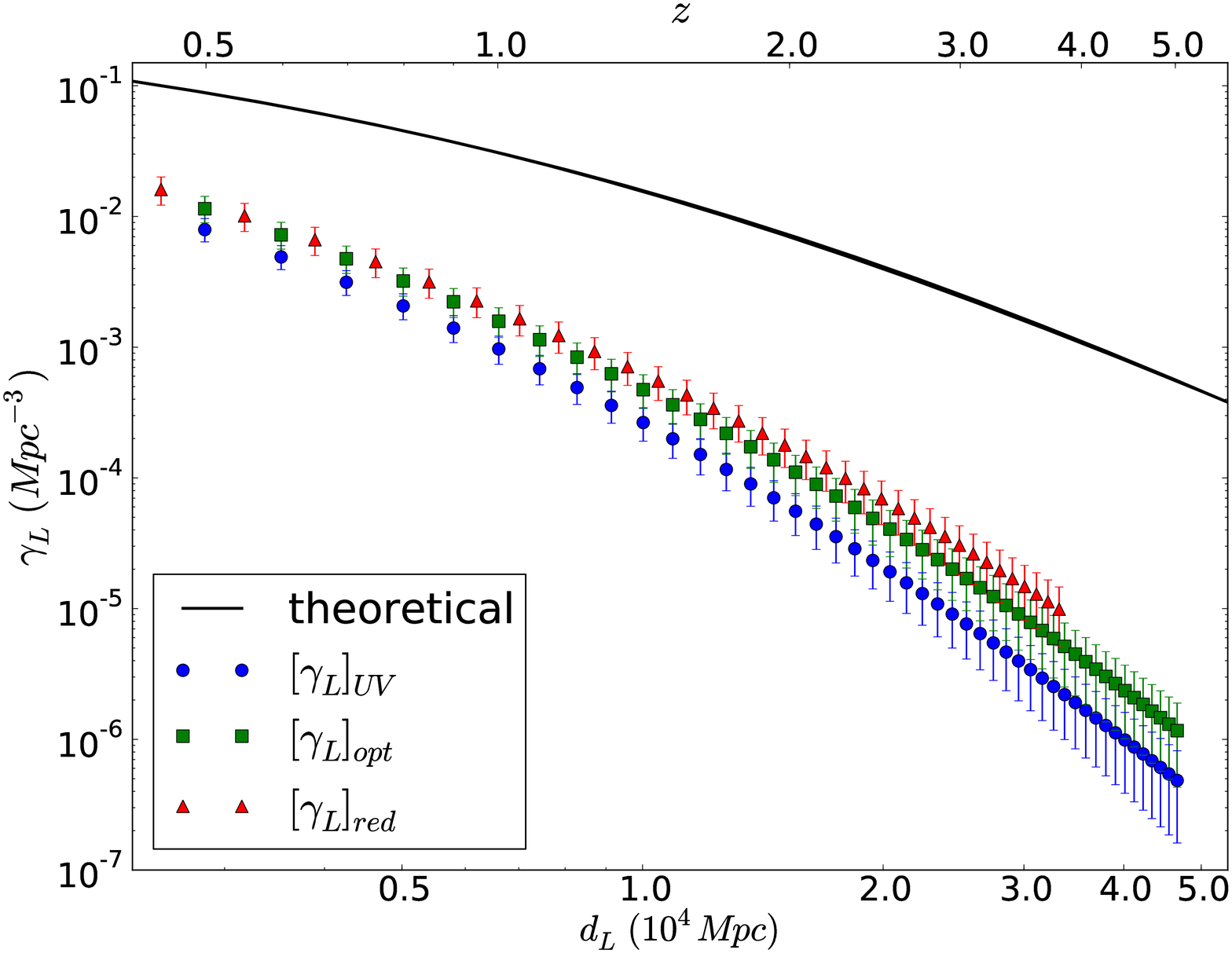}
&
\includegraphics[width=9.5cm]{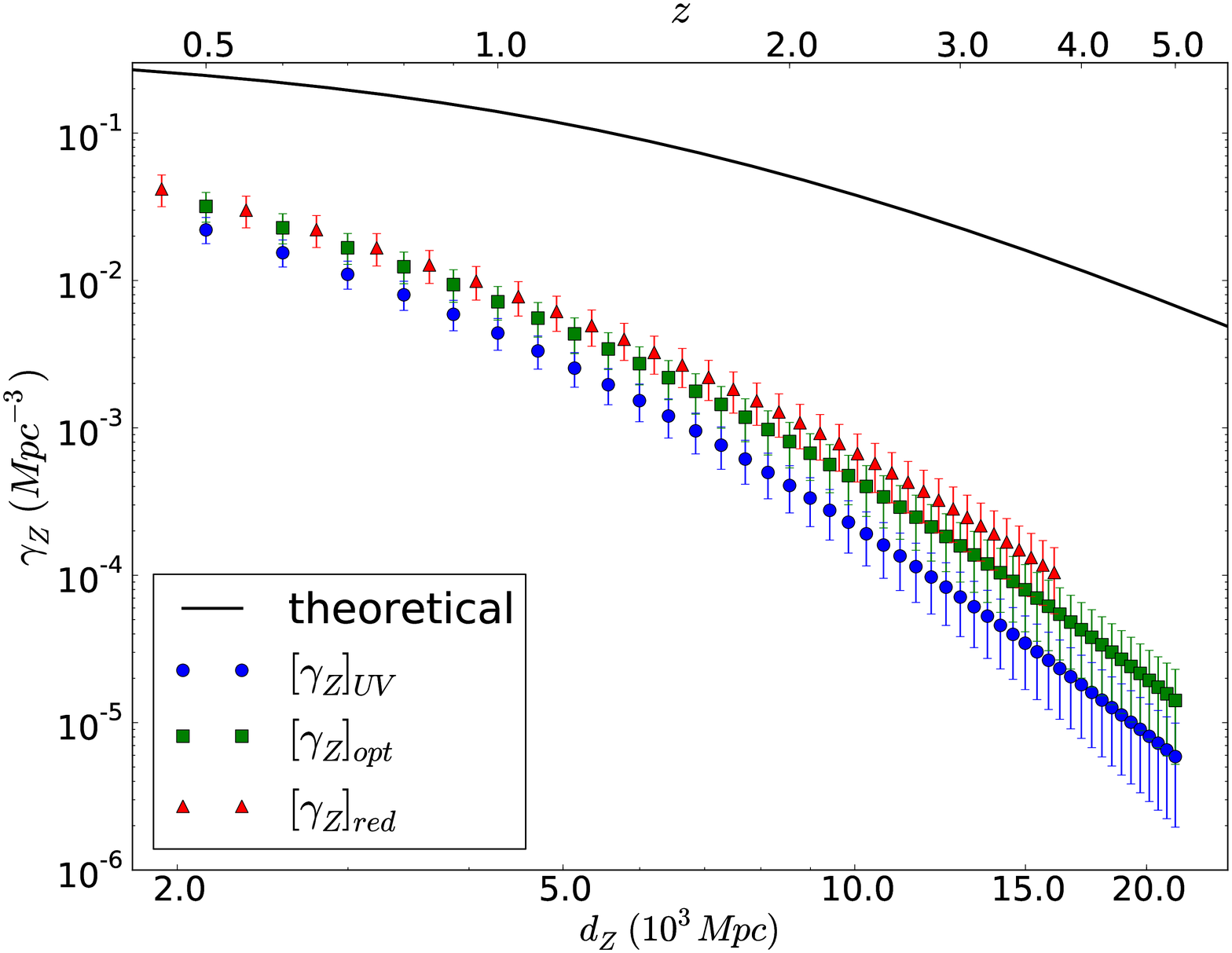}\\
\includegraphics[width=9.5cm]{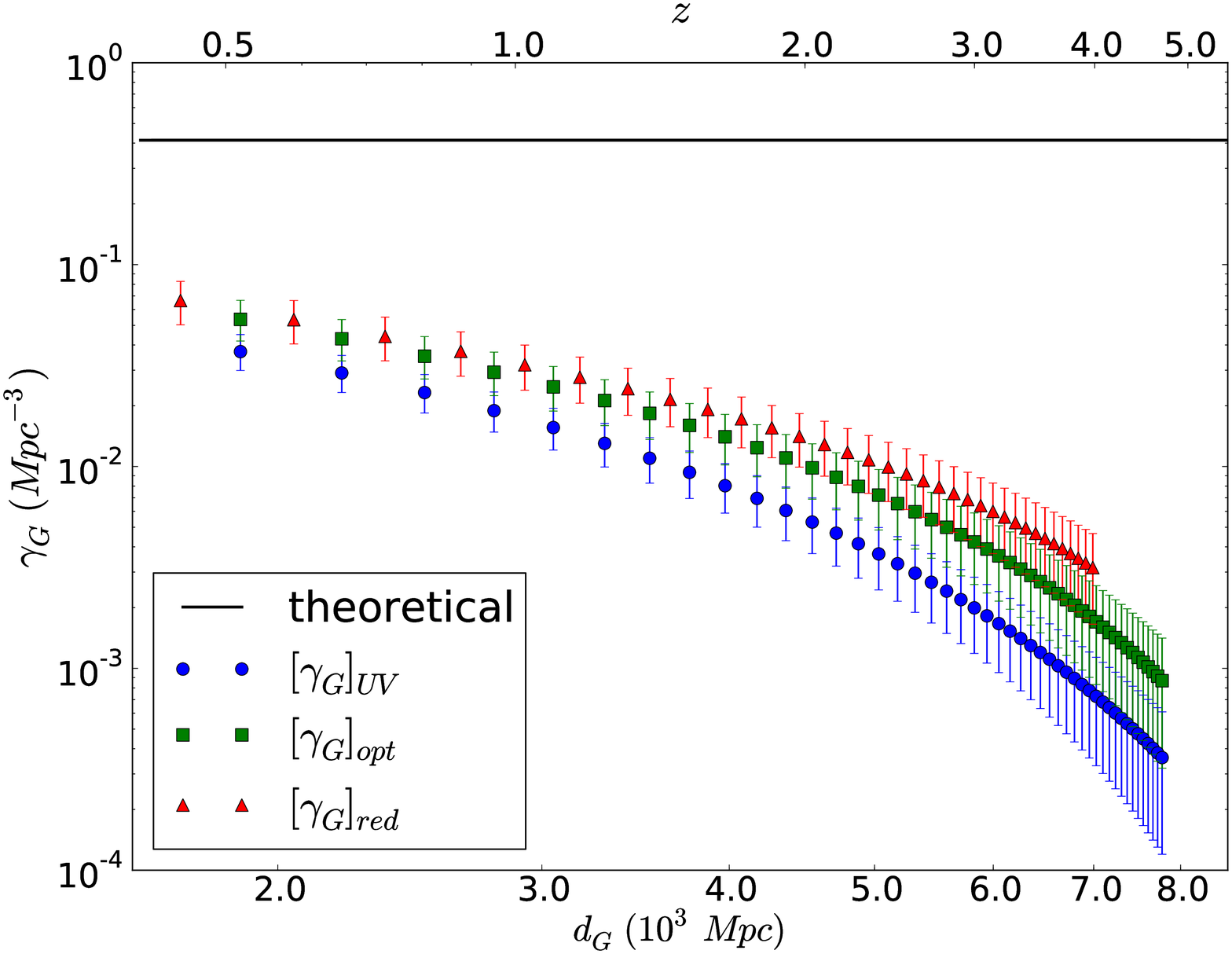}
&
\includegraphics[width=9.5cm]{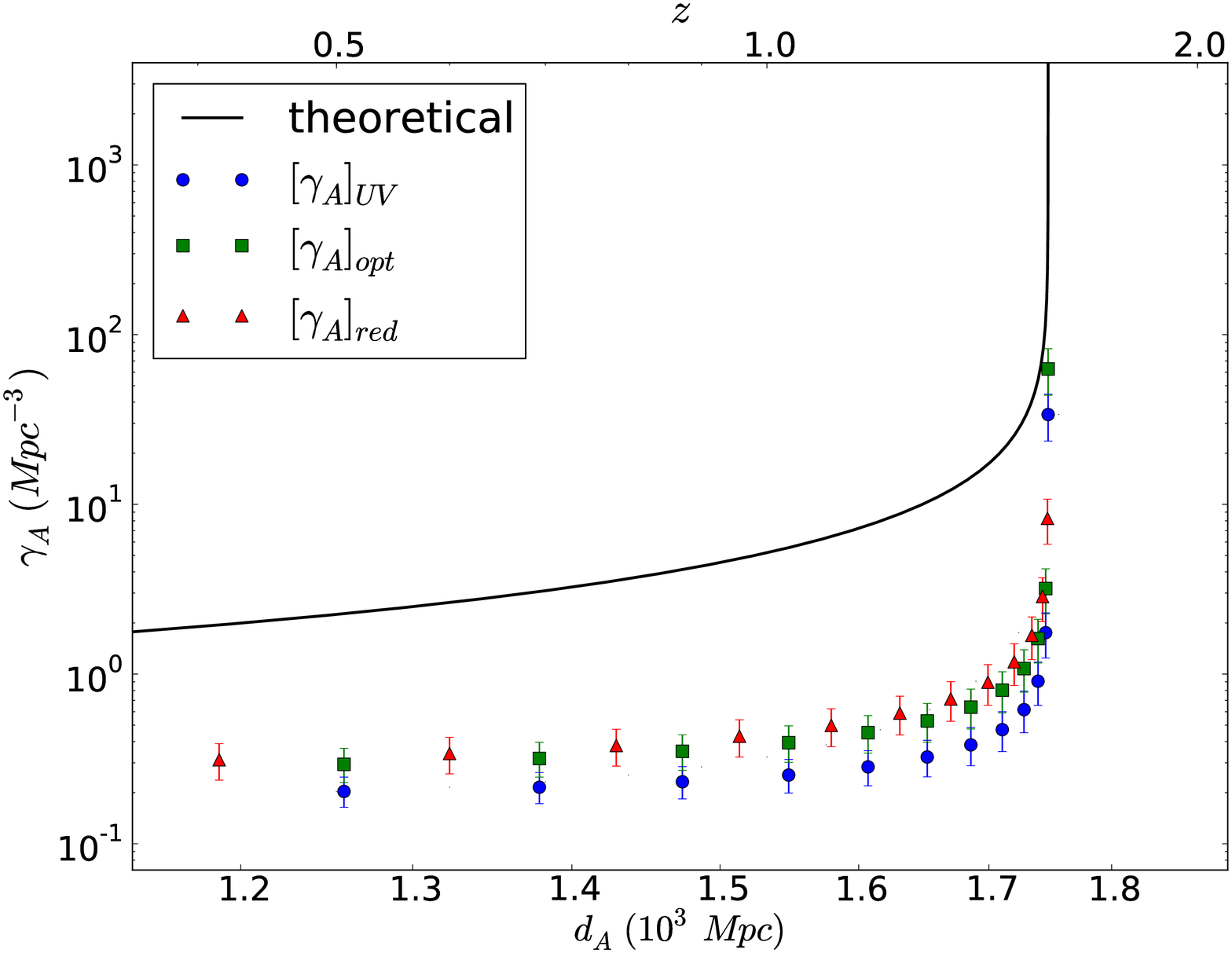}
\end{array}$
\end{center}
\caption{These plots show the theoretical and observational
relativistic differential densities for the four FLRW cosmological
distances adopted here. The observational densities are shown in
the combined UV, optical and red bands of the FDF datasets of G04
and G06. The difference between the theoretical curves and
observational ones comes from the fact that the SF is limited by
the observations themselves, which reflects on $J(z)$. So, such
difference indicates large quantities of dark matter, as already
pointed out by RS03 (see Sec.\ 5.1). Subsequent plots will also
reflect the same behavior. Symbols are as in the legend.}
\label{fig-gamma}
\end{figure*}

\begin{figure*}
\begin{center}$
\begin{array}{cc}
\includegraphics[width=9.5cm]{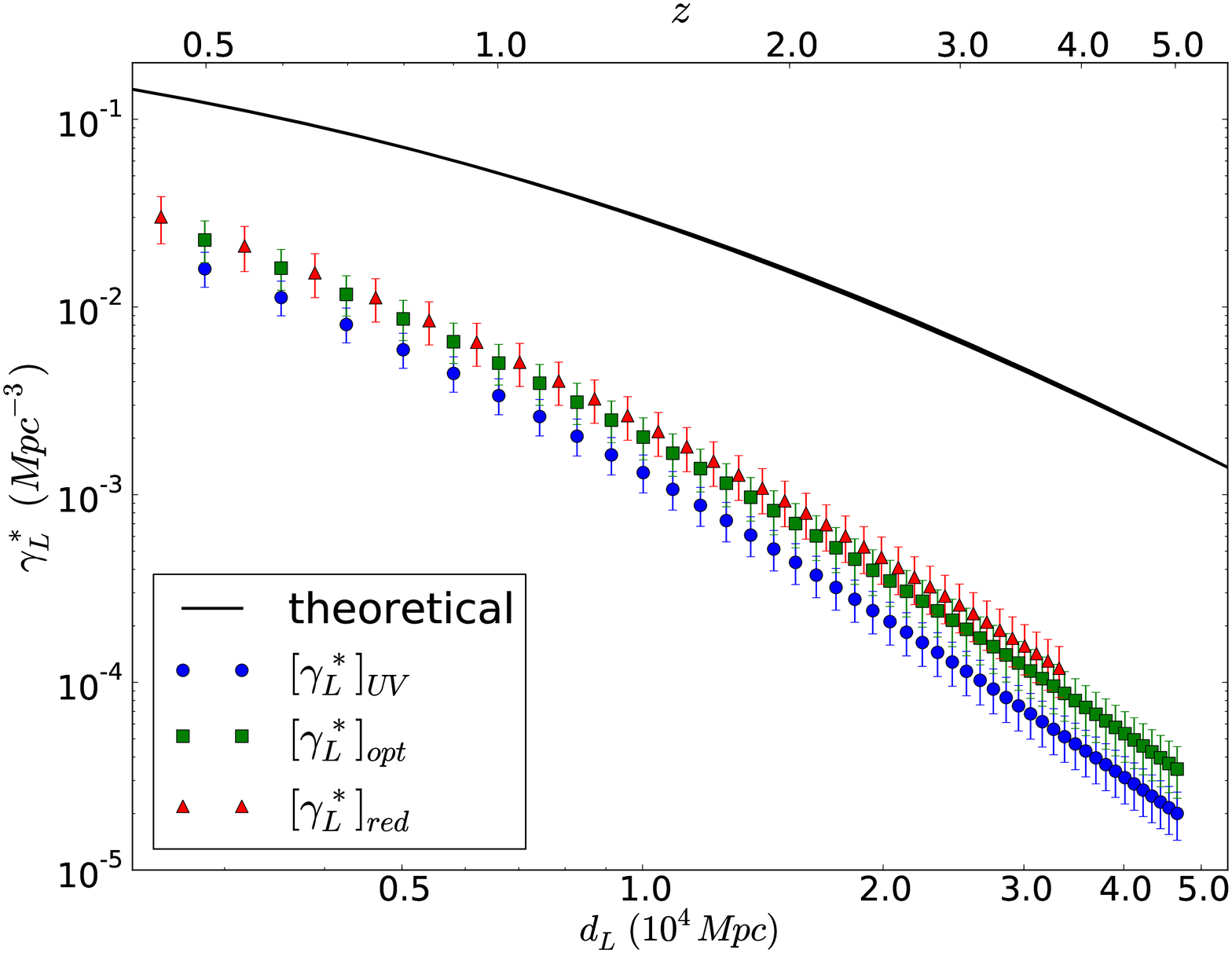}
&
\includegraphics[width=9.5cm]{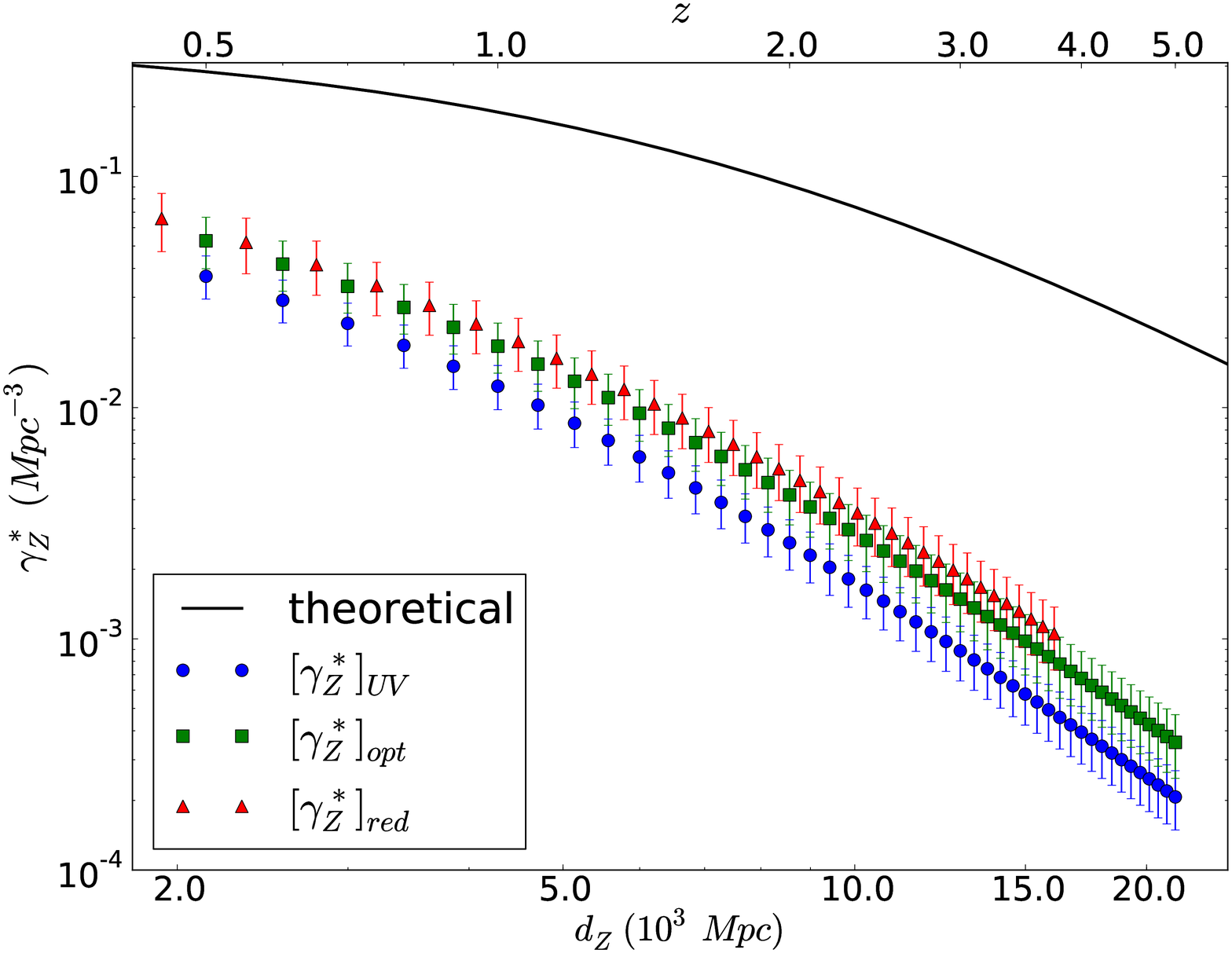}\\
\includegraphics[width=9.5cm]{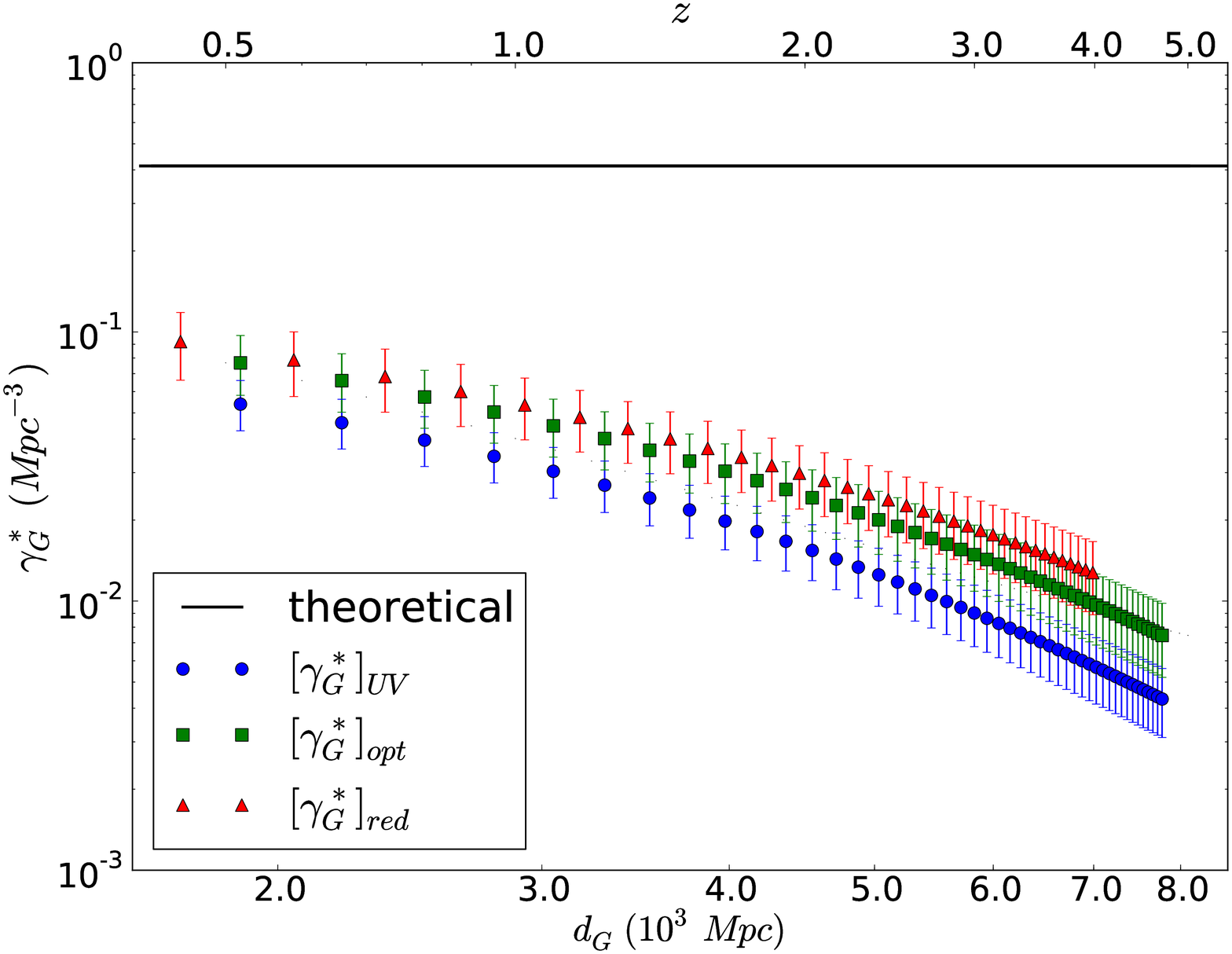}
&
\includegraphics[width=9.5cm]{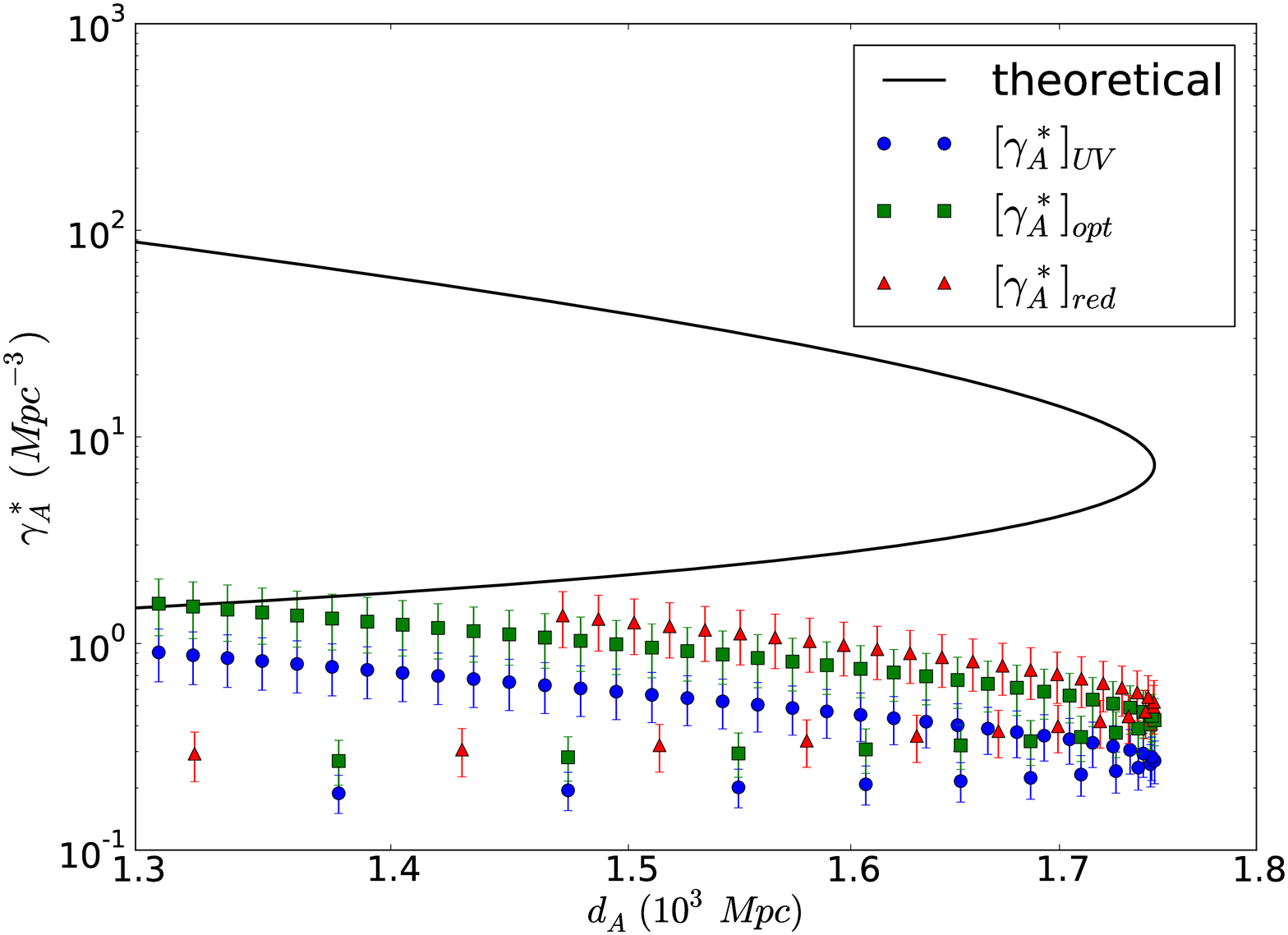}
\end{array}$
\end{center}
\caption{Plots showing the theoretical and observational relativistic
integral densities for the four FLRW cosmological distances adopted
here. The observational points are shown in the combined UV, optical
and red bands of the FDF datasets of G04 and G06. Symbols are as in
the legend.}
\label{fig-n}
\end{figure*}

The observational values of the differential number counts
$\obs{\drm N/\drm z}$ for all filters in the dataset allowed us
to evaluate the observational differential number densities
$\obs{\gamma}$ for the FLRW cosmology distance measures 
discussed above by simply replacing $\drm N/\drm z$ with
$\obs{\drm N/\drm z}$ in Eq.\ (\ref{gamma}). The dependence
of these densities on the distance definitions are shown in
Fig.\ \ref{fig-gamma}. Similarly, the observational values of
the integral densities $\obs{\gamma^\ast}$ were obtained by
replacing $N$ with $\obs{N}$ in Eq.\ (\ref{gest3}). The results
are shown in Fig.\ \ref{fig-n}.

\begin{figure*}
\begin{center}
\includegraphics[width=9.5cm]{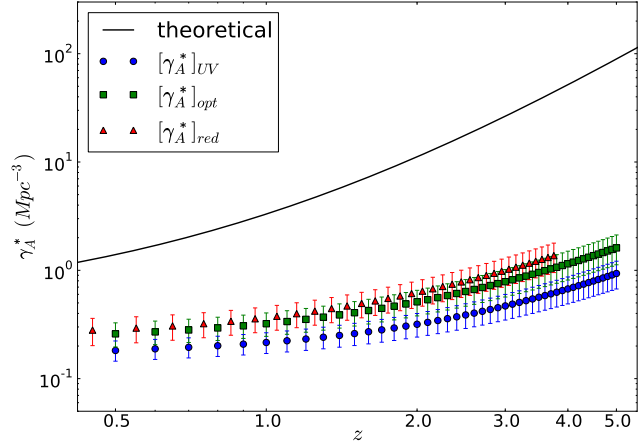}
\end{center}
\caption{This graph shows the theoretical and observational
relativistic integral densities obtained with the angular diameter
distance $\da$ against the redshift. The observational points are
in the combined UV, optical and red bands of the FDF datasets of
G04 and G06. Note the difference between this plot ($\gama^\ast$ vs.\
$z$) and the one in the lower right panel of Fig.\ \ref{fig-n}
($\gama^\ast$ vs.\ $\da$), although the Y-axis scale is basically
the same. Symbols are as in the legend.}
\label{fig.nA-z}
\end{figure*}

\begin{figure*}
\begin{center}
\includegraphics[width=18.2cm]{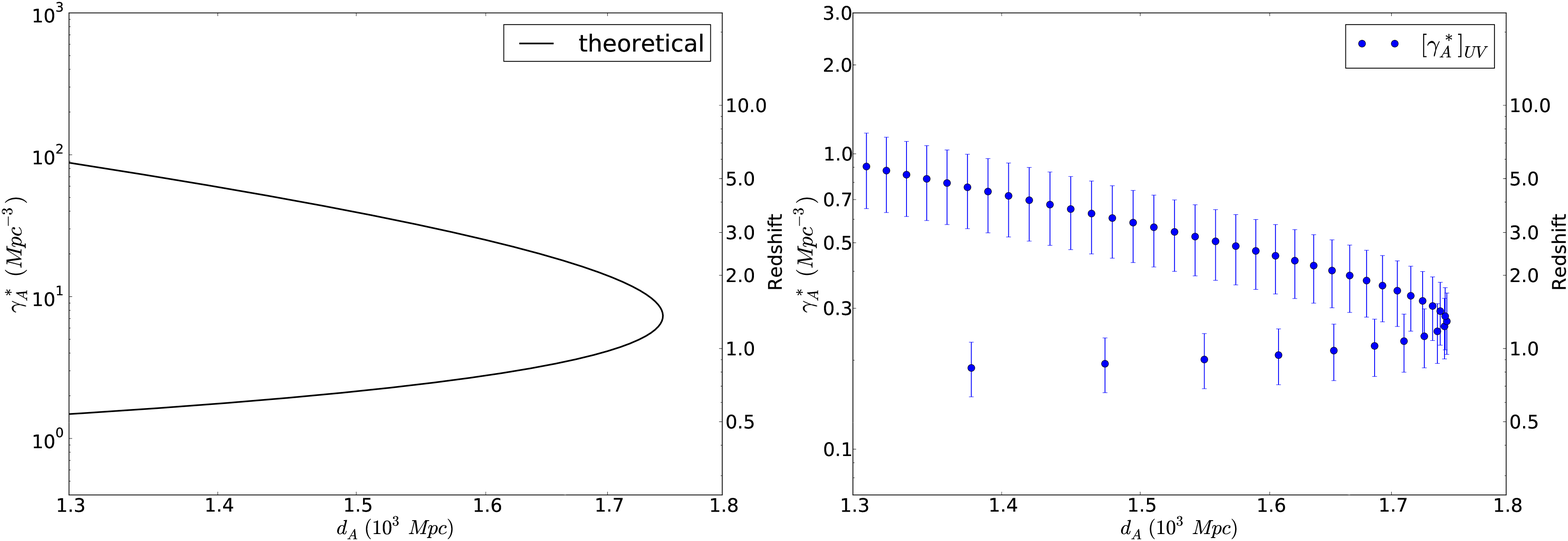}\\
\includegraphics[width=18.2cm]{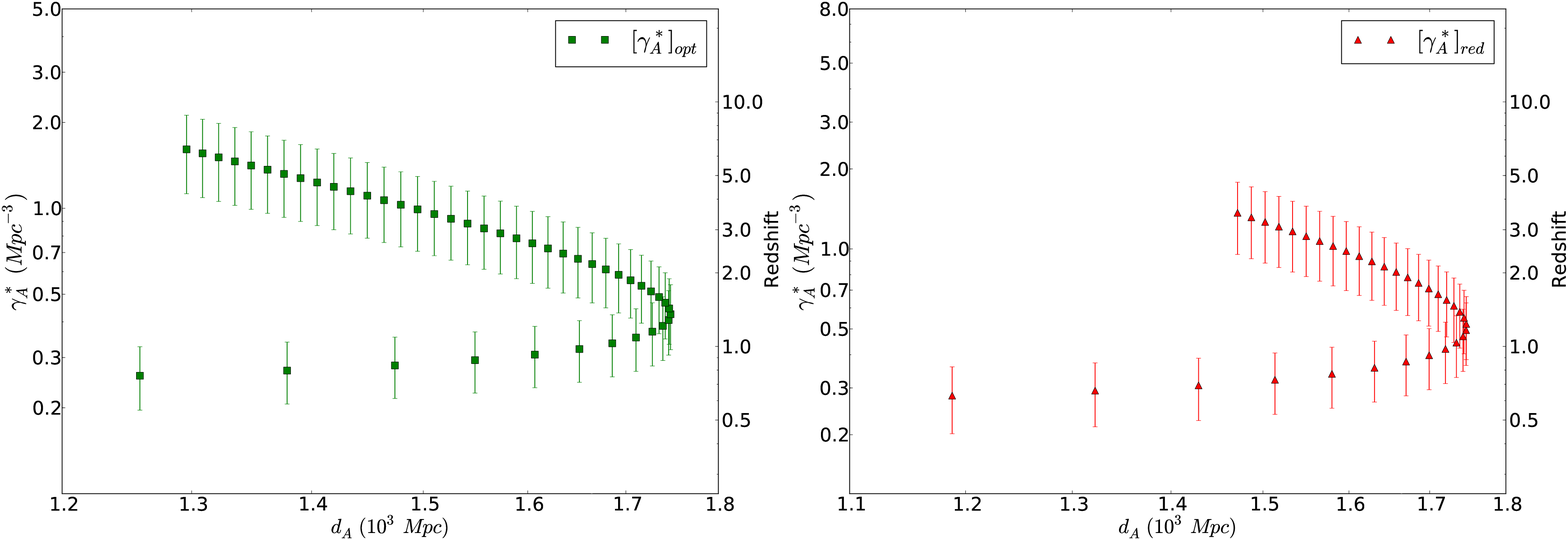}
\end{center}
\caption{This figure shows the theoretical and observational relativistic
integral densities computed with the angular diameter distance versus
$\da$ and the redshift. These graphs combine the information of Figs.\
\ref{fig.distances} and \ref{fig.nA-z}, basically splitting the lower
right plot of Fig.\ \ref{fig-n} in four different plots and adding a
redshift scale at the right side of each graph. Note that for the same
redshift scale on the right side of each plot one needs a different scale
for $\gama^{\ast}$ on the left side.}
\label{fig.nA-dA-z}
\end{figure*}

The differential density constructed with the area distance
$\da$ becomes discontinuous at $z \approx 1.6$ because $\da$
reaches a maximum at that redshift and, therefore, its derivative
with respect to $z$ vanishes, rendering  $\obs{\gamma_{\ssty A}}$
undefined. In addition, due to this maximum $\obs{\gamma^\ast_{\ssty
A}}$ has a curious turn around in $\da$ at $z \approx 1.6$, as shown
in Fig.\ \ref{fig-n}. Nevertheless, Fig.\ \ref{fig.nA-z} shows that
such a curious behavior does not occur when the integral density
$\gamma^\ast_{\ssty A}$ is plotted in terms of $z$, which in this
case keeps on increasing. This is due to the fact that each $\da$
corresponds to two redshift values (see Fig.~\ref{fig.distances}),
but since each $z$ produces only one $\gama^\ast$, each $\da$
generates then two values for $z$ and $\gama^\ast$. The single point
where this behavior is absent is when $\da$ reaches its maximum,
which corresponds to single values for $\da$, $z$ and $\gama^\ast$.
This effect is better visualized in Fig.~\ref{fig.nA-dA-z} where a
redshift scale was added to the right side of each plot.

\begin{figure*}
\begin{center}$
\begin{array}{cc}
\includegraphics[width=9.5cm]{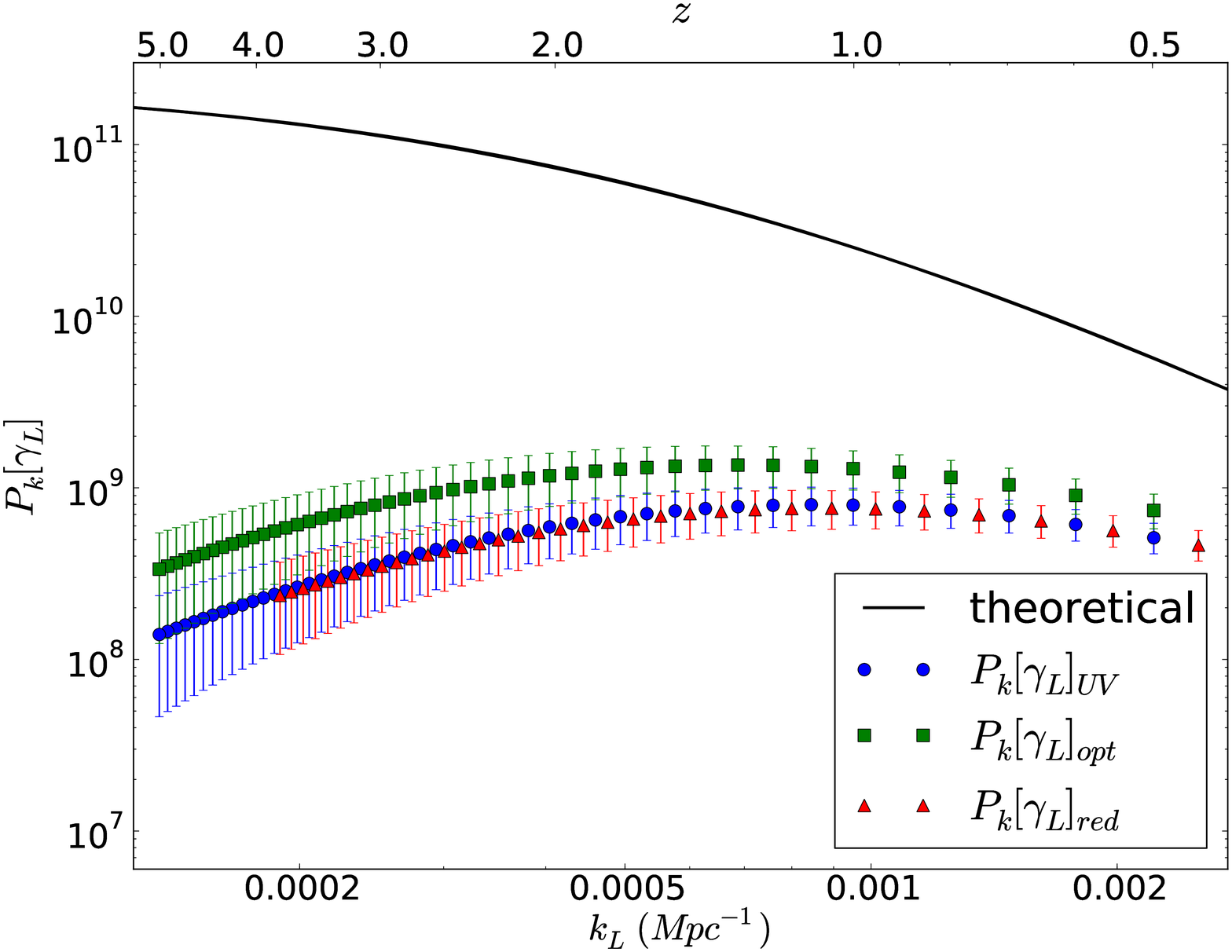}
&
\includegraphics[width=9.5cm]{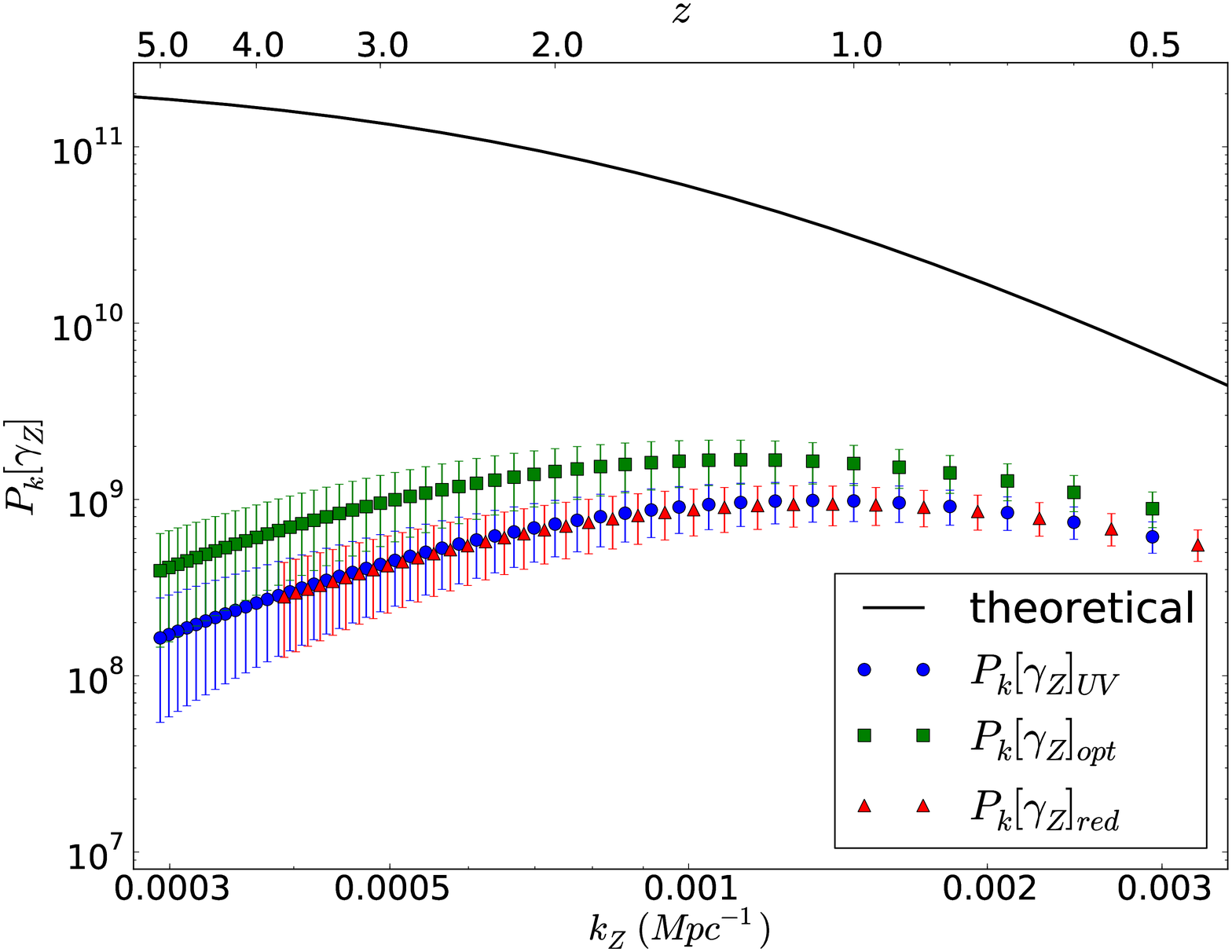}\\
\includegraphics[width=9.5cm]{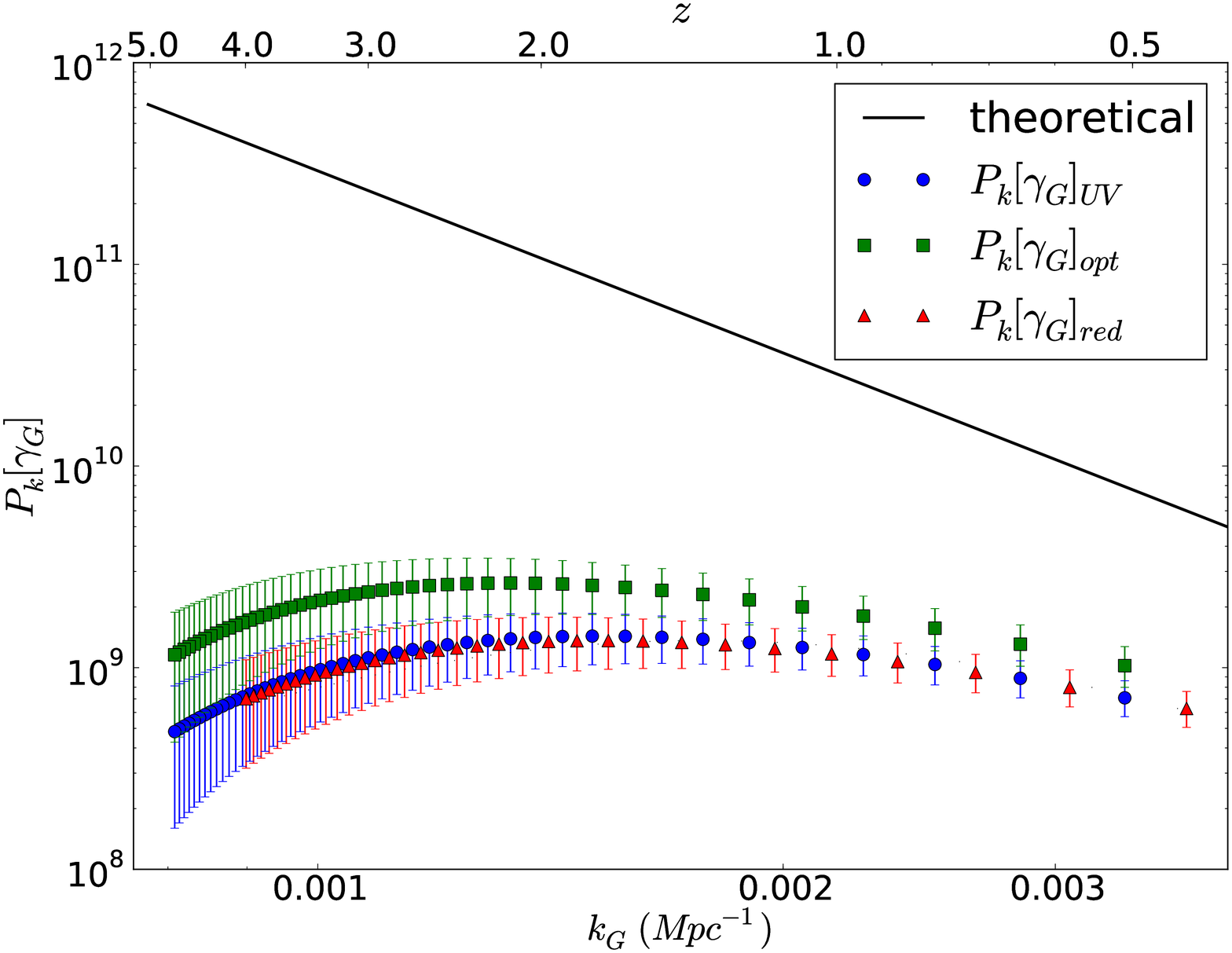}
&
\includegraphics[width=9.5cm]{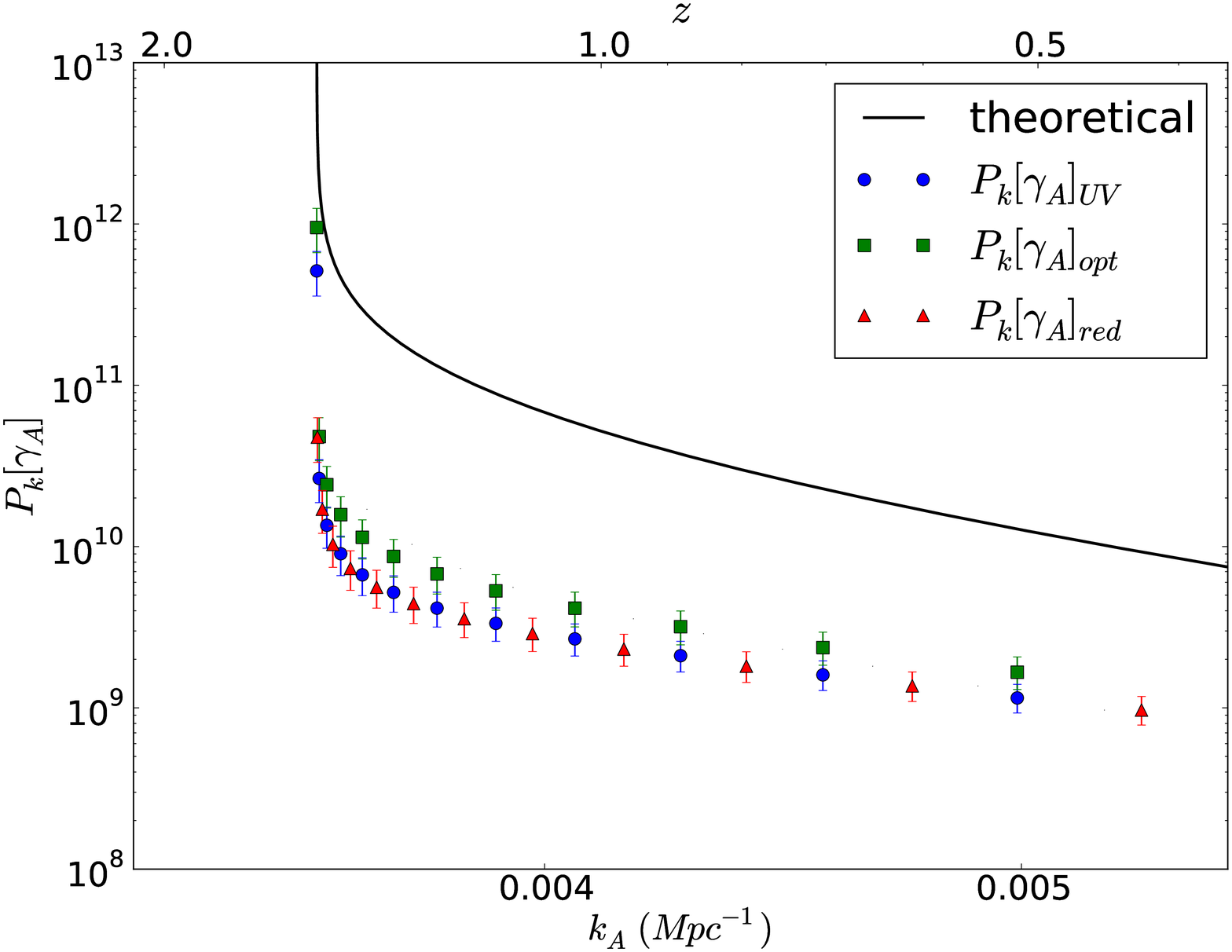}
\end{array}$
\end{center}
\caption{This figure shows a panel with the power spectra of the
theoretical and observational relativistic differential densities
computed with the four FLRW cosmological distances adopted here.
The observational points are presented in the combined UV, optical
and red bands of the FDF datasets of G04 and G06. Symbols are as
in the legend.}
\label{fig.PS-gamma}
\end{figure*}

Fig.\ \ref{fig.PS-gamma} shows the PS of the FDF observational
differential densities ${P_k}\obs{\gamma_i}$ and their theoretical
counterparts plotted against the wave number $k_i$ and the
redshift. The observational points were determined by means of the
combined UV, optical and red bands in the G04 and G06 dataset. It
is clear from the graphs that $P_k\obs{\gama}$ possesses an odd
behavior due to its discontinuity at $z \approx 1.6$, bearing no
resemblance to the PS of the two-point correlation function,
whereas the other power spectra $P_k\obs{\gamma_{i={\ssty G, \;
L, \; Z}}}$ show a similar behavior to the two-point correlation
function PS as presented by Tegmark et al.\ (2004, Fig.\ 38). We
note that these authors calculated the two-point correlation
function PS using different types of data, such as cosmic
background radiation, galaxies, clusters of galaxies, lensing
and Lyman alpha forest. Nevertheless, their general PS behavior
is basically similar to ours, apart from the one calculated with
$\da$.

\begin{figure*}
\begin{center}$
\begin{array}{cc}
\includegraphics[width=9.5cm]{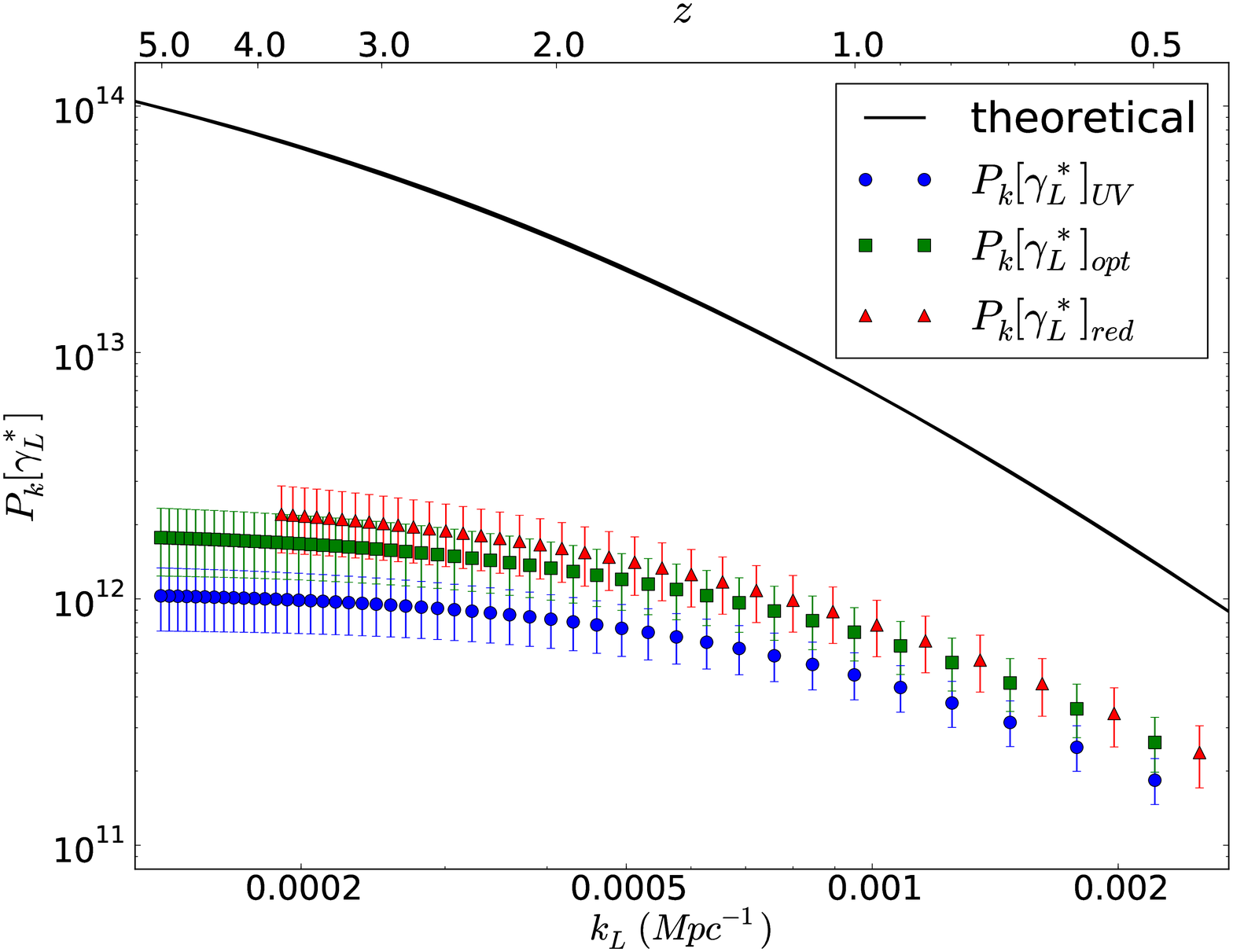}
&
\includegraphics[width=9.5cm]{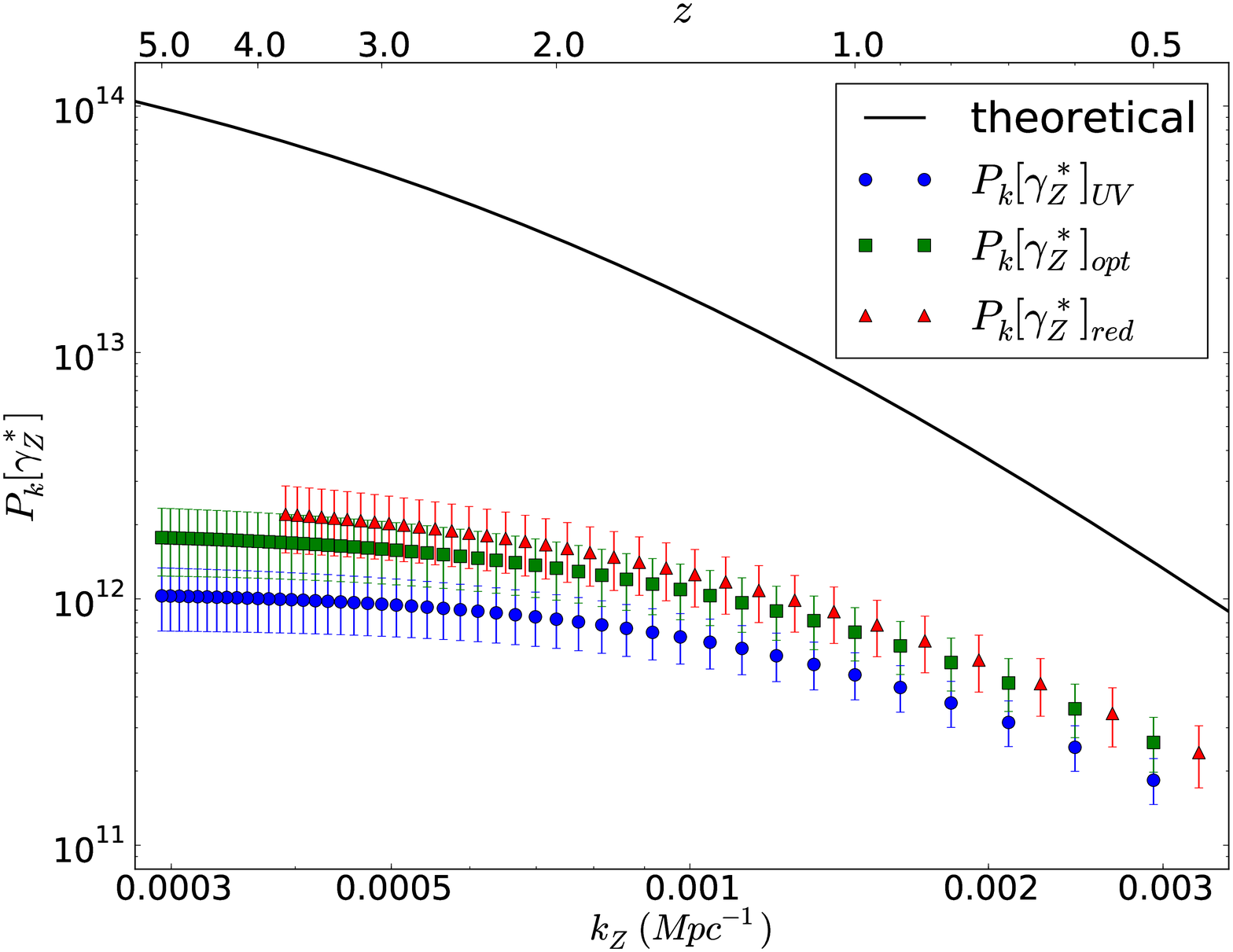}\\
\includegraphics[width=9.5cm]{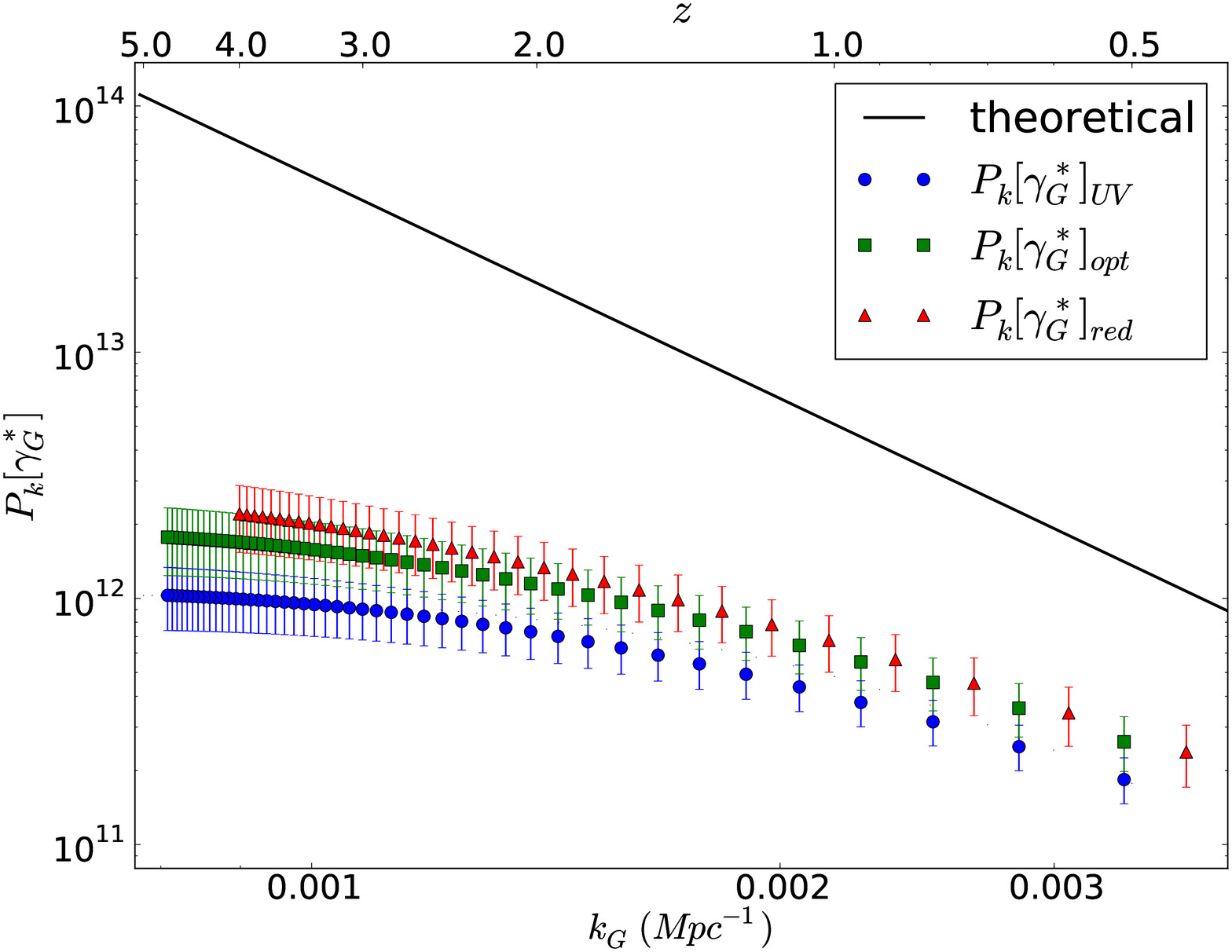}
&
\includegraphics[width=9.5cm]{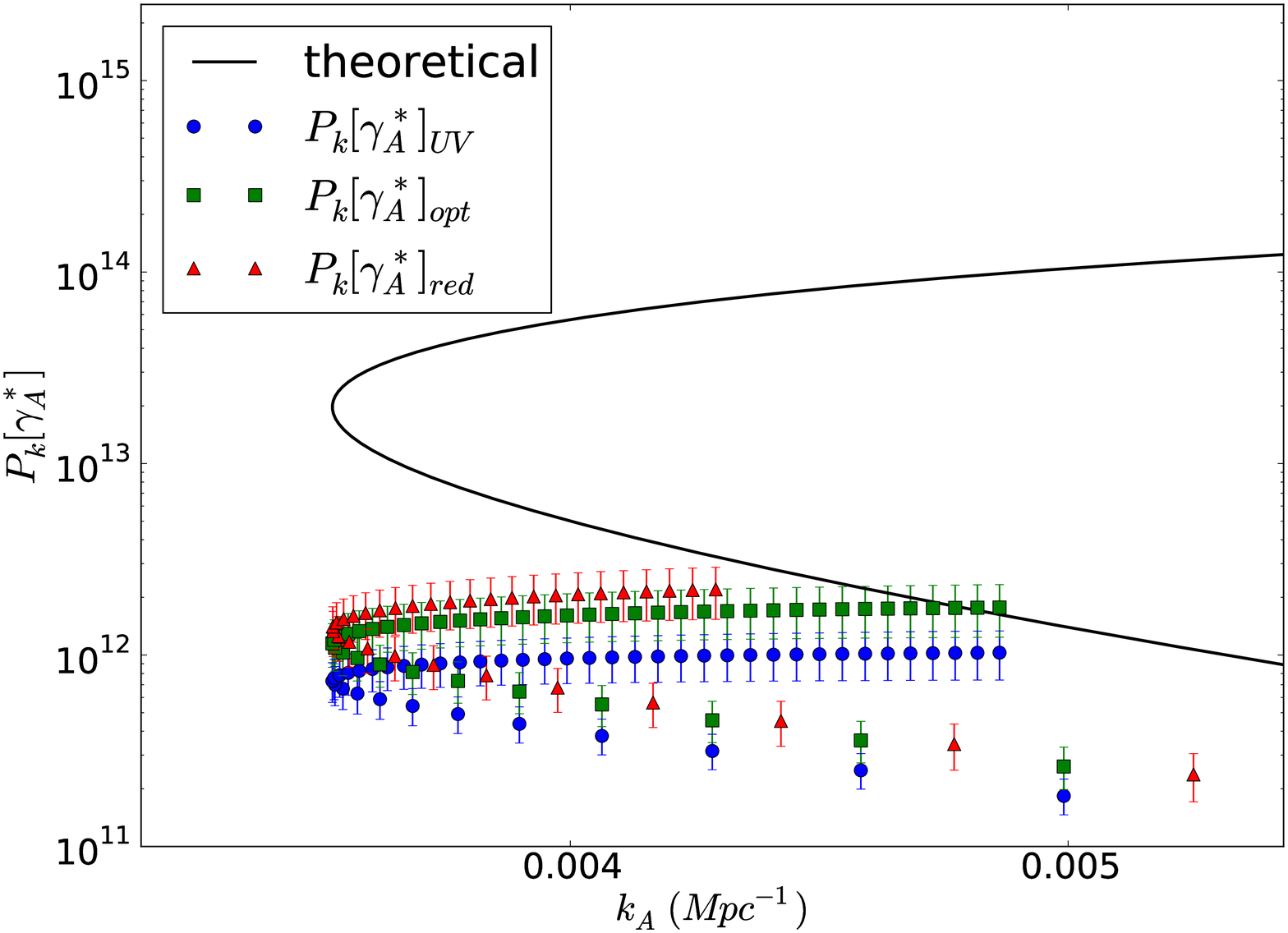}
\end{array}$
\end{center}
\caption{Panel showing the power spectrum of theoretical and observational
relativistic integral densities. The observational points are presented in 
the combined UV, optical and red bands of the FDF datasets of G04 and G06.
Symbols are as in the legend.}
\label{fig.PS-n}
\end{figure*}

The PS of the observational integral densities in the UV, optical
and red combined bands versus the wave number and the redshift
are shown in Fig.\ \ref{fig.PS-n}. We note that
$P_k\bigobs{\gamma^\ast_{i={\ssty G, \; L, \; Z}}}$ differ from from
$P_k\obs{\gamma_{i={\ssty G, \; L, \; Z}}}$ shown in the previous
figure only at small wave number values or, equivalently, at
higher redshifts. Such difference occurs because the differential
densities measure the rate of growth in number counts since
$\gamma_i \propto \df N/ \df z$, whereas $\gamma^\ast_i \propto N$
(see Eqs.\ \ref{gamma} and \ref{gest3}). As $N$ is a cumulative
quantity, it can only increase or remain constant, while $\df N/
\df z$ increases, reaches a maximum and then decreases at scales
dependent on the ratio of increase in $N$. Clearly the declining
behavior of $\df N/ \df z$ will occur at higher values of $z$,
beyond its maximum, leading the decline in $\gamma$ to become
even more pronounced at those scales. In addition, by measuring
the rate of growth in number counts, $\gamma$ is more sensitive
to local fluctuations due to noisy data. Thus, the steeper decline
detected in the slopes of $P_k\obs{\gamma_{i={\ssty G, \; L, \; Z}}}$
at small wave numbers as compared to those at
$P_k\bigobs{\gamma^\ast_{i={\ssty G, \; L, \; Z}}}$ can be attributed
to these distortion effects at the redshift limits of the sample.

\begin{figure*}
\begin{center}
\includegraphics[width=9.5cm]{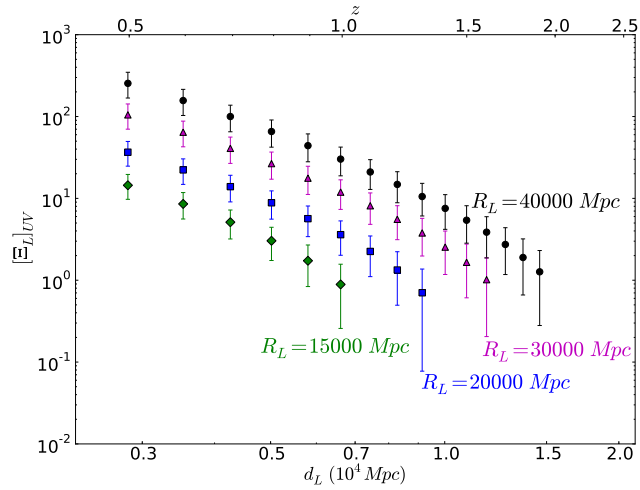}
\end{center}
\caption{Graph of the radial correlation for the luminosity distance in the
combined UV waveband for different sample sizes. Symbols are as in the legend.}
\label{fig.RadialCorr_size}
\end{figure*}

\begin{figure*}
\begin{center}$
\begin{array}{cc}
\includegraphics[width=9.5cm]{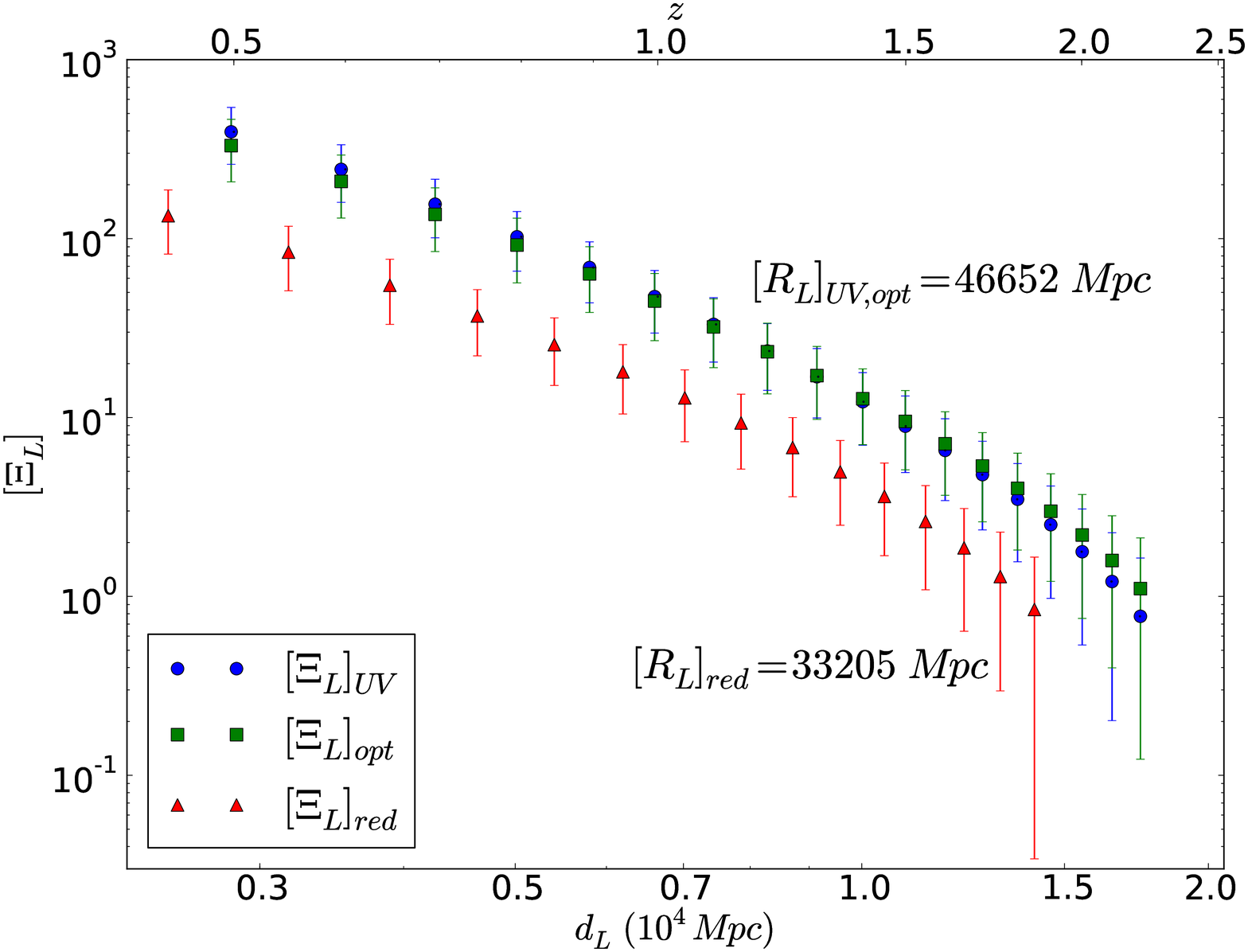}
&
\includegraphics[width=9.5cm]{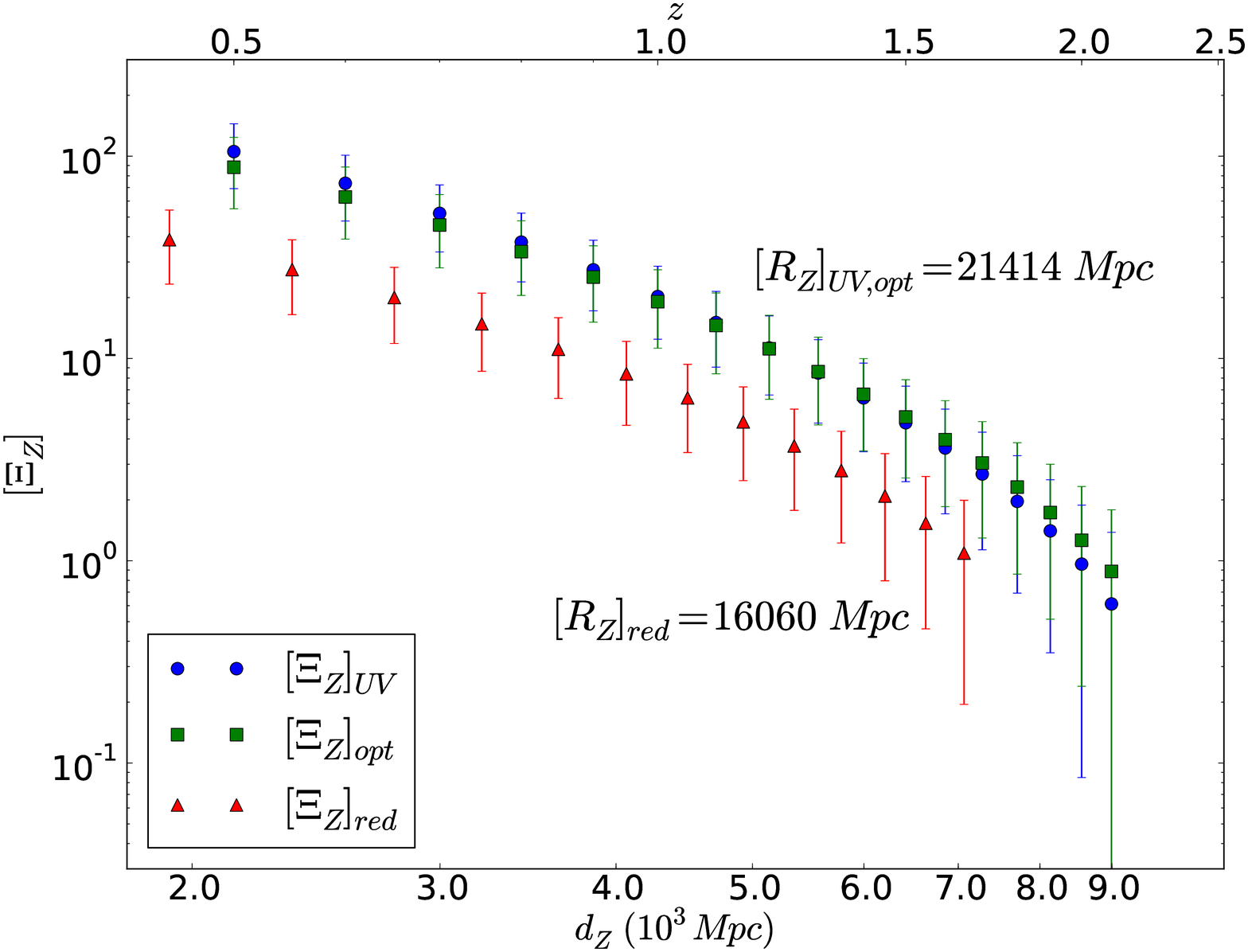}\\
\includegraphics[width=9.5cm]{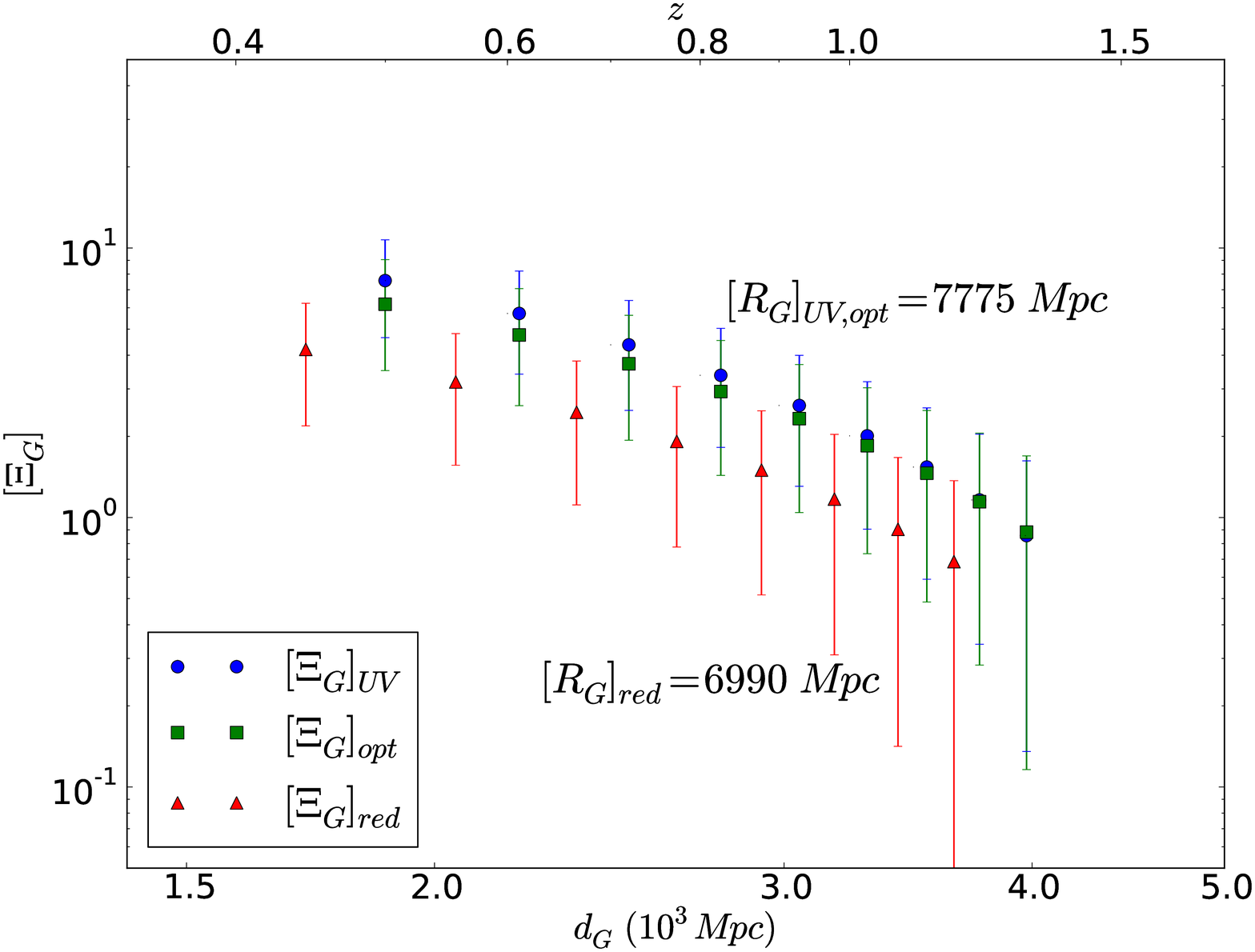}
\end{array}$
\end{center}
\caption{Radial correlation for the luminosity distance, redshift
	distance and galaxy area distance in the combined UV, optical
	and red bands of the FDF datasets of G04 and G06. Symbols are
        as in the legend.}
\label{fig.RadialCorr}
\end{figure*}

As discussed above, the radial correlation $\Xi_i (d_i)$ proposed by
Ribeiro (1995) was then studied in the Einstein-de Sitter model, but
only with the luminosity distance. Here we extend the analysis of
Eq.\ (\ref{radial-corr}) to the galaxy area distance and redshift
distance in the FLRW model with $\Omega_{m}=0.3$ and $\Omega_{\Lambda}
=0.7$. These results are shown in Fig.\ \ref{fig.RadialCorr_size}
where one can see that, similarly to the Einstein-de Sitter cosmology,
the amplitude of $\obs{\Xi_{\ssty L}}$ in the FLRW model also depends
on the sample size. The extension of these results to $\dz$ and $\dg$
are presented in Fig.\ \ref{fig.RadialCorr}, where different sample
sizes were used, confirming thus such a dependence. Due to the
pathological behavior of $\da$, $\obs{\Xi_{\ssty A}}$ was not evaluated.

\begin{figure*}
\begin{center}$
\begin{array}{cc}
\includegraphics[width=9.5cm]{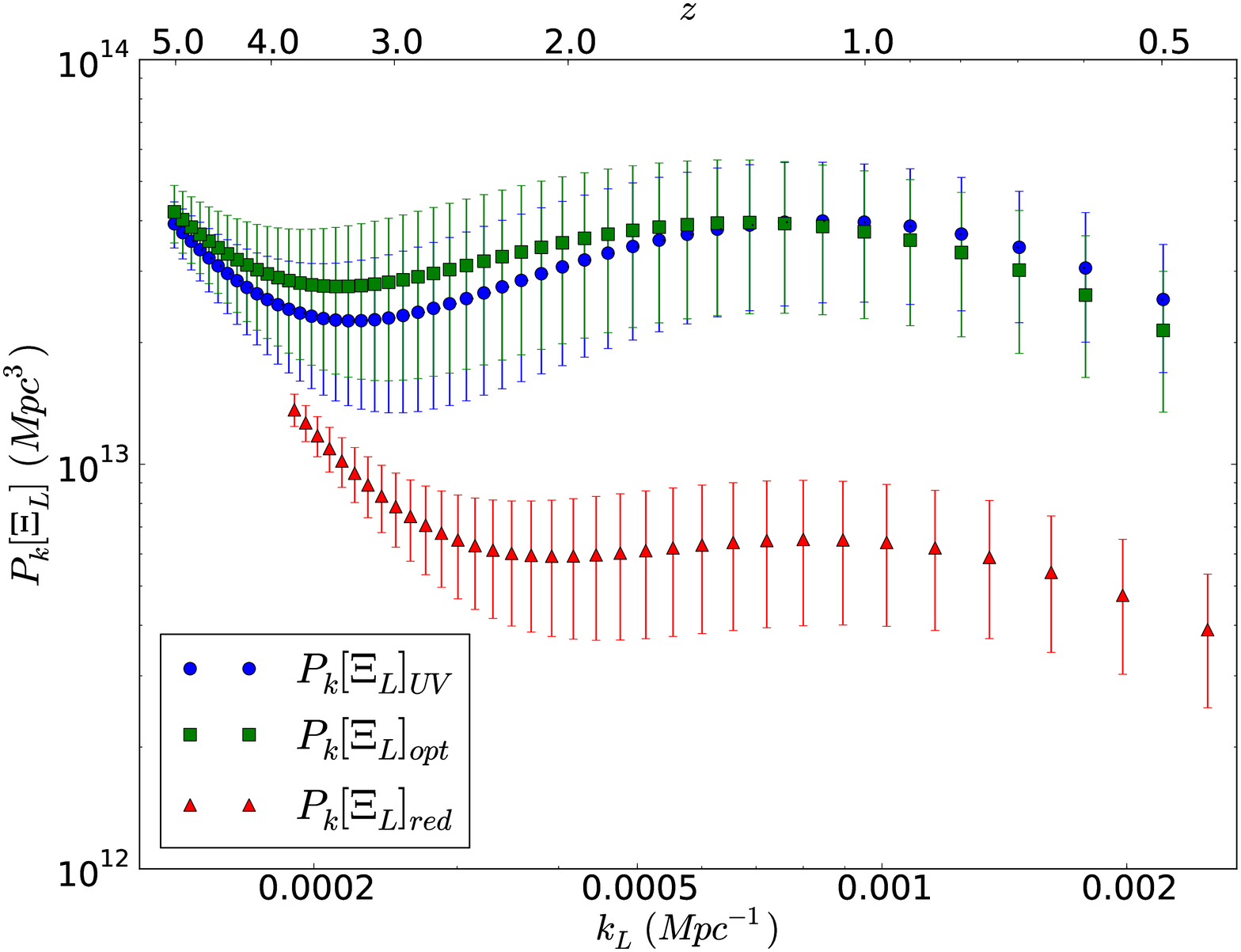}
&
\includegraphics[width=9.5cm]{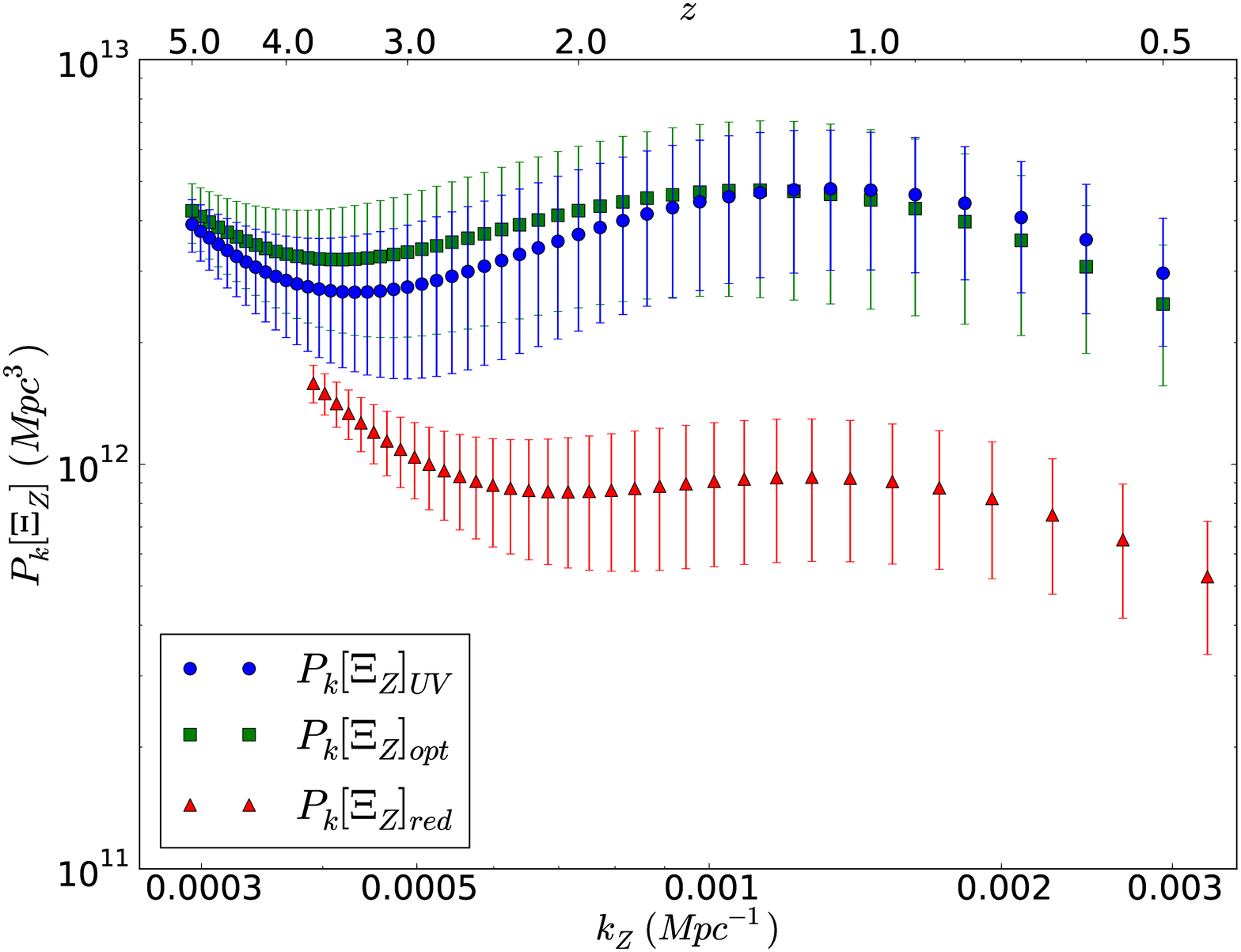}\\
\includegraphics[width=9.5cm]{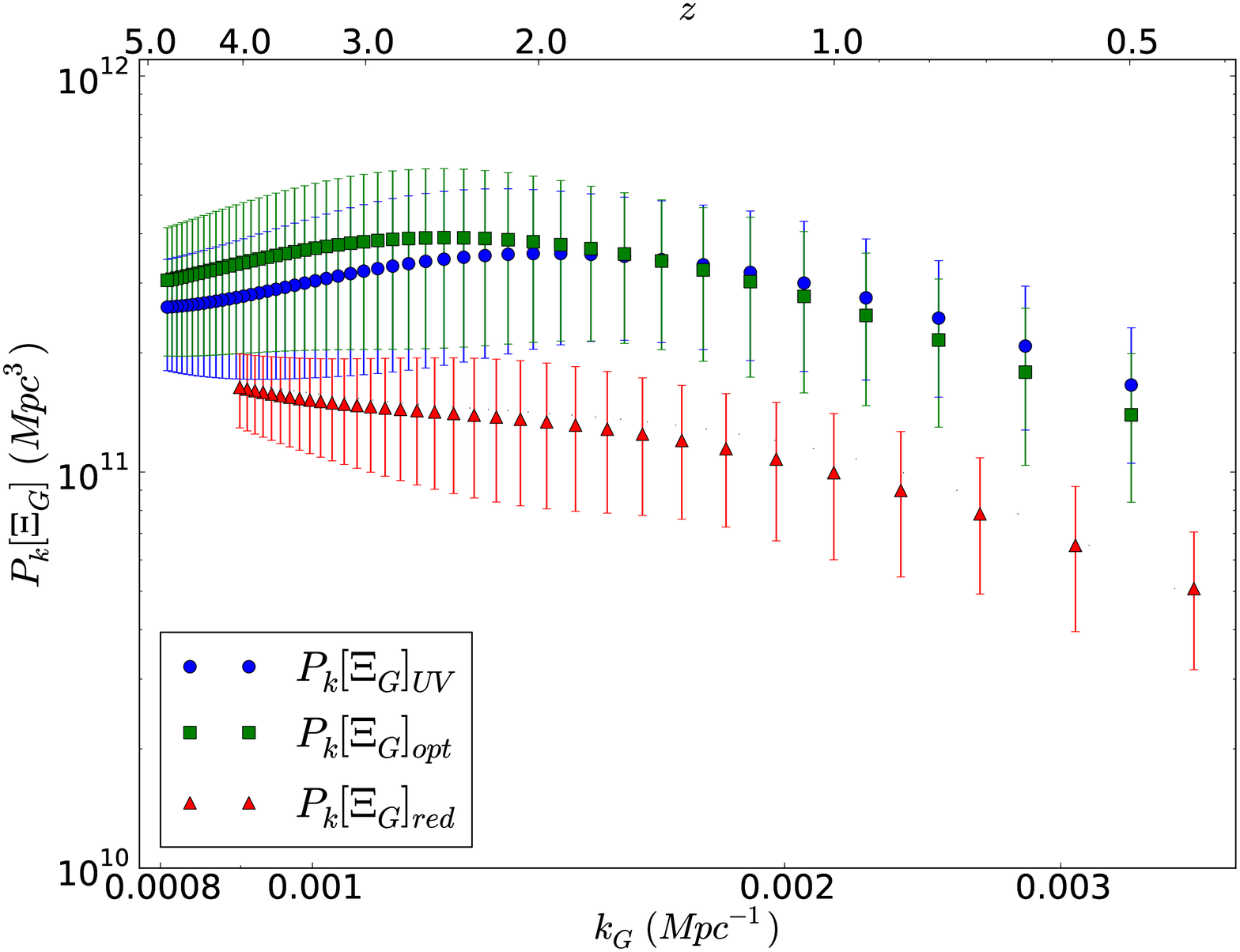}
\end{array}$
\end{center}
\caption{PS of Radial correlation for the luminosity distance, redshift
	distance and galaxy area distance in the combined UV, optical and
	red bands of the FDF datasets of G04 and G06. Values of $\obs{R_i}$
        are as in Fig.\ \protect\ref{fig.RadialCorr}. Symbols are as in the
        legend.}
\label{fig.PS-RadialCorr}
\end{figure*}

\begin{figure*}
\begin{center}$
\begin{array}{cc}
\includegraphics[width=9.5cm]{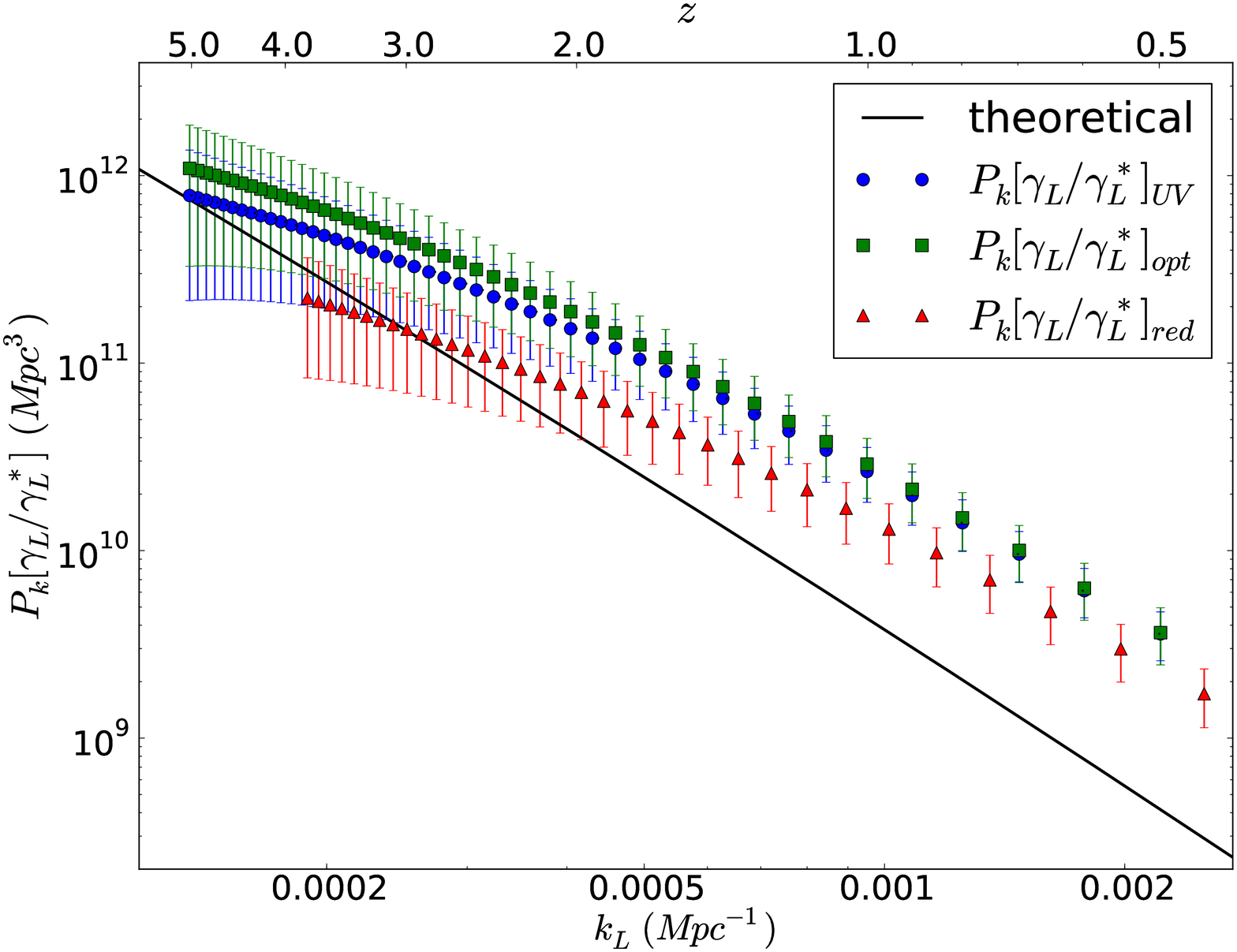}
&
\includegraphics[width=9.5cm]{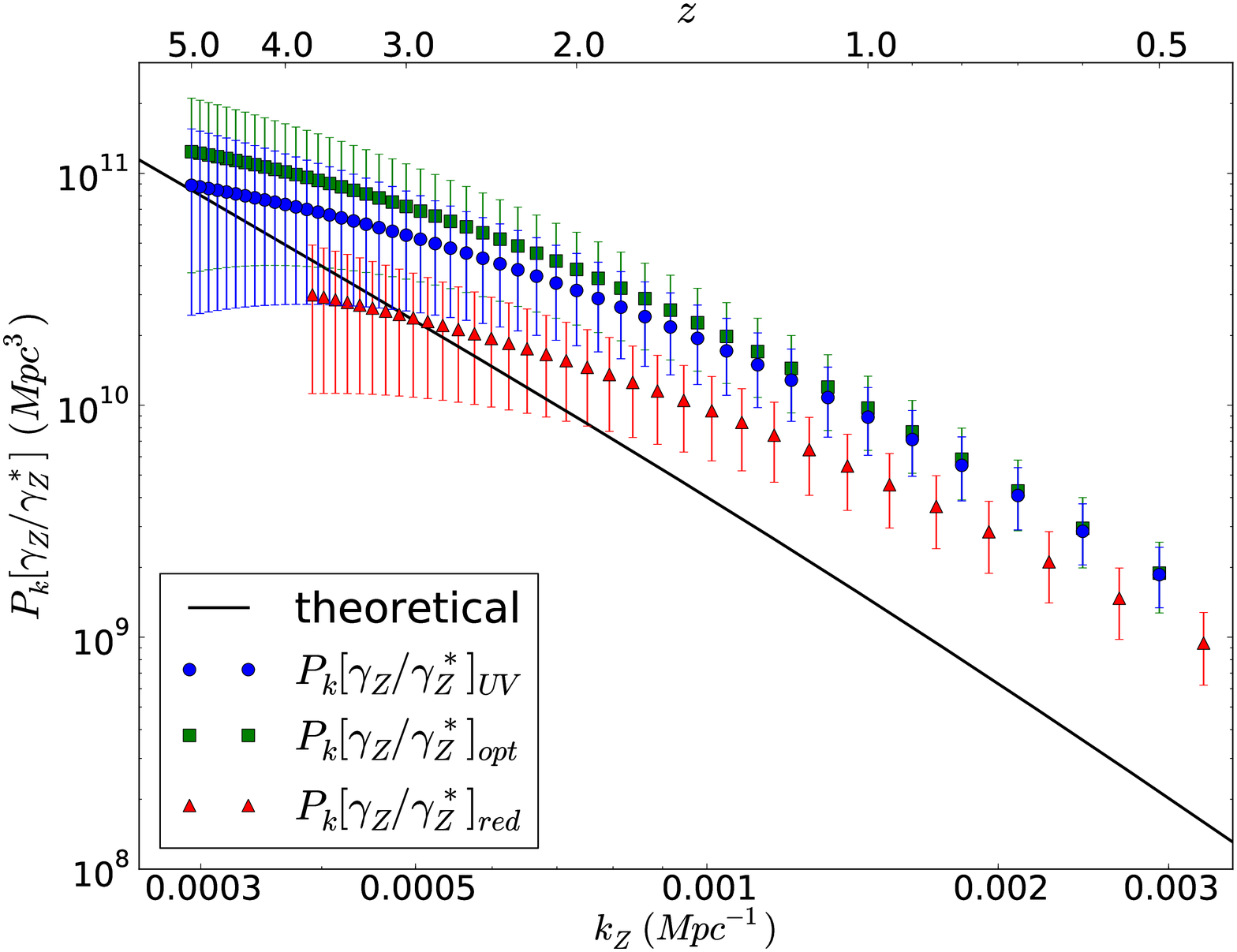}\\
\includegraphics[width=9.5cm]{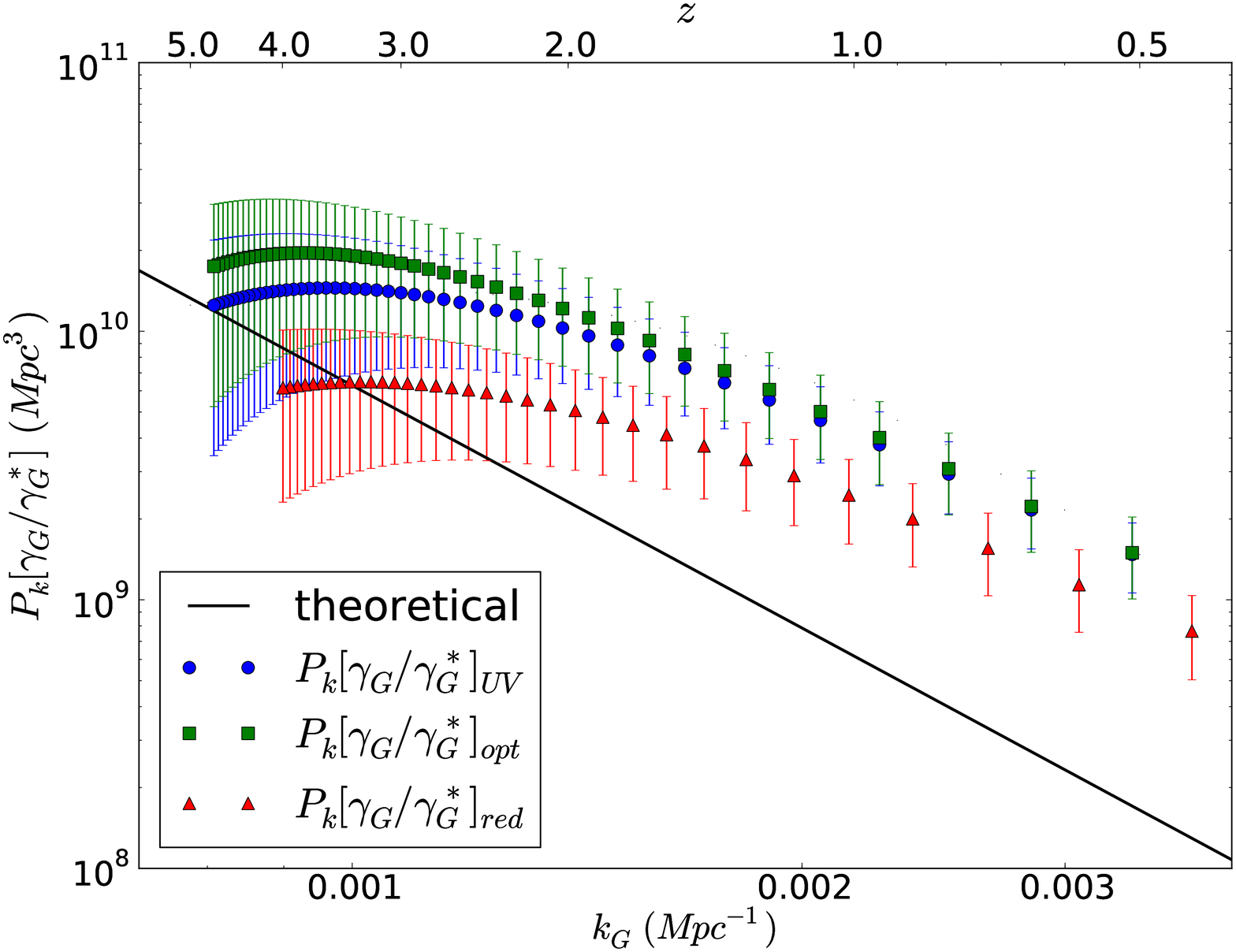}
\end{array}$
\end{center}
\caption{PS of the ratio $\gamma / \gamma^{*}$ for the luminosity distance,
	redshift distance and galaxy area distance in the combined UV,
	optical and red bands of the FDF datasets of G04 and G06. Values
	of $\obs{R_i}$ are as in Fig.\ \protect\ref{fig.RadialCorr}.
        Symbols are as in the legend.}
\label{fig.PS-Ratio}
\end{figure*}

Once in possession of the $\obs{\Xi_{i={\ssty G, \; L, \; Z}}}$
computations for the FDF dataset, it is natural to look at the
behavior of their respective power spectrum. Fig.\
\ref{fig.PS-RadialCorr} shows such PS results, where one can see a
distortion at small wave numbers probably caused by the subtraction
of one in Eq.\ (\ref{radial-corr}). To see if that is actually the
reason behind this distortion, we evaluated the PS of the ratio
$\gamma / \gamma^{*}$ for $\dl$, $\dz$ and $\dg$. Graphs of these
quantities are shown in Fig.\ \ref{fig.PS-Ratio} where we can
clearly see the absence of the previous distortion. This means that
as far the observed PS for the radial correlation $P_k\obs{\Xi_i}$
is concerned, this is better calculated if one redefines $\Xi_i$ in
Eq.\ (\ref{radial-corr}) as follows, 
\begin{equation}
P_k\obs{{\Xi_{i}(d_i)}} =
P_k\bigobs{\frac{{\gamma_{i}(d_i)}}{\gamma^{*}_{i}(R_{i})}}.
\label{radial-corr2}
\end{equation}

\begin{figure*}
\begin{center}
\includegraphics[width=17.4cm]{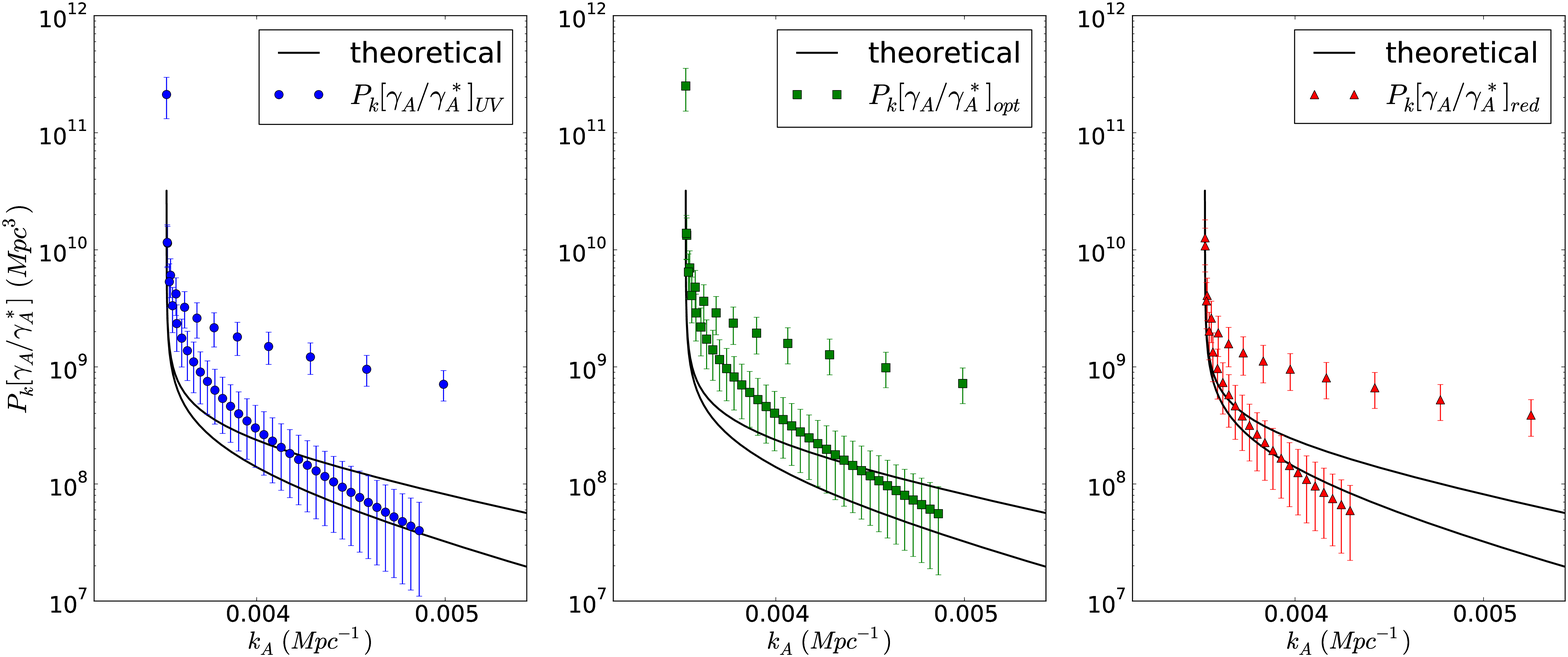}
\end{center}
\caption{PS of the ratio $\gamma / \gamma^{*}$ for the area distance in
	the combined UV, optical and red bands of the FDF datasets of G04
	and G06. Symbols are as in the legend.}
\label{fig.PS-RatioA}
\end{figure*}

Finally, the PS of the ratio $\gamma / \gamma^{*}$ for the area distance
are shown in Fig.\ \ref{fig.PS-RatioA} where the discontinuity in
$\gamma_{\ssty A}$ and $\obs{\gamma_{\ssty A}}$ translates themselves
to discontinuous behavior for $P_k \obs{\gamma_{\ssty A} /
\gamma_{\ssty A}^\ast}$. 

\section{Conclusion}\lb{conclusions}

In this paper we have studied the theoretical and observational
power spectrum of the differential and integral galaxy volume
number densities $\gamma$ and $\gamma^\ast$ using various
relativistic cosmology distance measures $d_i$, namely the
angular diameter distance $\da$, galaxy area distance $\dg$,
luminosity distance $\dl$ and redshift distance $\dz$ $(i =
{\sty A},{\sty G},{\sty L}\, ,{\sty Z})$. The cosmology
adopted here is the FLRW model with non zero cosmological
constant $\Lambda$ having $\Omega_{m_{\ssty 0}}=0.3$,
$\Omega_{\Lambda_{\ssty 0}} = 0.7$ and $H_0=70 \; \mbox{
km} \; {\mbox{s}}^{-1} \; {\mbox{Mpc}}^{-1}$. Observational
results were obtained by following the framework that connects
the relativistic cosmology number counts theory with the
astronomical data extracted from the galaxy luminosity
function (LF), as proposed by Ribeiro \& Stoeger (2003) and
further developed by Albani \etal (2007) and Iribarrem \etal
(2012). This framework uses the LF parameters of the FDF
galaxy survey provided by Gabasch \etal (2004, 2006) in order
to calculate the observed differential number counts
$\obs{\drm N/\drm z}$ required to build the observational
densities $\obs{\gamma_i}$ and $\obs{\gamma^\ast_i}$. The
observational results were produced in the combined red, blue
optical and UV bands in the redshift range $0.5 \le z \le 5.0$. 

The PS of the galaxy volume number densities were obtained directly
from both $\obs{\gamma_i}$ and $\obs{\gamma^\ast_i}$ and two other 
combinations of these quantities, namely the ratio $\obs{\gamma_i /
\gamma^\ast_i}$ and the radial correlation $\obs{\Xi_i}$ which
mixes up these two densities in a way that they can be discussed in
terms of the sample size and each distance measure used above.
Therefore, our analysis was able to compute $P_k \obs{\gamma_i}$,
$P_k \obs{\gamma_i^\ast}$, $P_k \obs{\gamma_i / \gamma_i^\ast}$ and
$P_k \obs{\Xi_i}$. 

The results show that the graphs for $P_k \obs{\gamma_{i={\ssty G,
\; L, \; Z}}}$, $P_k \bigobs{\gamma_{i={\ssty G, \; L, \; Z}}^\ast}$
and $P_k \bigobs{{( \gamma_i / \gamma_i^\ast )}_{i={\ssty G, \; L,
\; Z}}}$ have similar behavior for larger wave numbers, but the first
differs from the other two at small wave numbers. Both $P_k
\bigobs{\gamma_{i={\ssty G, \; L, \; Z}}^\ast}$ and $P_k \bigobs{{(
\gamma_i / \gamma_i^\ast )}_{i={\ssty G, \; L, \; Z}}}$ have general
power-law behavior, whereas $P_k \obs{\gamma_{i={\ssty G, \; L, \; Z}}}$
behaves more similarly to the PS derived from the 2-point correlation
function (see Fig.\ 38 of Tegmark et al.\ 2004) having a more pronounced
decline at very small wave numbers. 

The PS analysis using $\da$ is problematic because this distance
measure starts increasing, reaches a maximum and then decreases,
resulting in a discontinuous $\obs{\gamma_{\ssty A}}$ at $z
\approx 1.6$. Therefore, both $P_k \obs{\gamma_{\ssty A}}$ and
$P_k \bigobs{{ \gamma_{\ssty A} / \gamma_{\ssty A}^\ast }}$ also
present discontinuities. $P_k \obs{\gamma^\ast_{\ssty A}}$ is,
however, continuous, but has a minimum wave number value just as
$\da$ has a maximum.

The radial correlation ${\Xi_i}$, a quantity defined to possibly
exhibit sample dependence in observational values, actually
showed such a dependence for $\obs{\Xi_{i={\ssty G, \; L, \; Z}}}$,
confirming a previous theoretical prediction made by using only the
luminosity distance $\dl$ (Ribeiro 1995). Nevertheless, their
respective PS $P_k \bigobs{\Xi_{i={\ssty G, \; L, \; Z}}}$ present
some strong distortions.

In summary, the study presented here suggests that the PS analysis
of the galaxy number densities can be considered as a complementary
tools for studies of the large-scale galaxy distribution. That may be
particularly the case of models describing the galaxy distribution as
a fractal system, as advanced by Conde-Saavedra et al.\ (2015) and
Teles et al.\ (2021), because these works showed that
both $\gamma_i$ and $\gamma_i^\ast$ are essential tools for the
fractal characterization of the galaxy distribution. This is particularly
the case because these densities present decaying power-law behavior for
$i=({\sty G},{\sty L},{\sty Z})$ at increasing distances, as shown in
Figs.\ \ref{fig-gamma} and \ref{fig-n}. Since their respetive power
spectra also behave mostly as decaying power-laws for larger wave numbers,
as seen in Figs.\ \ref{fig.PS-gamma} and \ref{fig.PS-n}, as well as with
the radial correlations in Fig.\ \ref{fig.RadialCorr} and their redefined
power spectra (Eq.\ \ref{radial-corr2}) as shown in Fig.\ \ref{fig.PS-Ratio},
these quantities provide additional tools for probing the large-scale galaxy
distribution in the universe.

\section*{Data Availability Statement}
All data generated or analysed during this study are included in this
published article.

\begin{acknowledgements}
We thank Filipe B.\ Abdalla for the original suggestion which
motivated this paper, and a referee for useful suggestions and
comments. Throughout this research A.R.L.\ was funded by
\textit{ Coordena\c{c}\~{a}o de Aperfei\c{c}oamento de Pessoal de
N\'{\i}vel Superior - Brasil (CAPES)} - Finance Code 001, to whom she is
grateful.

\end{acknowledgements}

\begin{thebibliography}{}
\bibitem{albani}{Albani, V.V.L., Iribarrem, A.S., Ribeiro, M.B.,
        and Stoeger, W.R.\ 2007, ApJ, 657, 760-772,
        arXiv:astro-ph/0611032 (\textbf{A07})}
\bibitem{gabi}{Conde-Saavedra, G., Iribarrem, A., Ribeiro, M.B.\ 2015,
        Physica A, 417, 332-344, arXiv:1409.5409}
\bibitem{ellis71}{Ellis, G.F.R.\  1971, in General Relativity
        and Cosmology, ed.\ R.K.\ Sachs (Proc.\ Int.\ School
	Phys.\ ``Enrico Fermi''; New York: Academic Press); reprinted
	in Gen.\ Rel.\ Grav., 41, 581, 2009}
\bibitem{ellis07}{Ellis, G.F.R.\ 2007, Gen.\ Rel.\ Grav., 39, 1047-1052}
\bibitem{etherington}{Etherington, I.M.H.\ 1933, Phil.\ Mag.\
         ser.\ 7, 15, 761; reprinted in Gen.\ Rel.\ Grav., 39, 1055, 2007}
\bibitem{FORS-blue}{Gabasch, A.\ \etal 2004, \aap, 421, 41-58 (\textbf{G04})}
\bibitem{FORS-red}{Gabasch, A.\ \etal 2006, \aap, 448, 101 (\textbf{G06})}
\bibitem{gabrielli05}{Gabrielli, A.\ \etal 2005, Statistical Physics for
        Cosmic Structures (Springer Verlag)}
\bibitem{heidt}{Heidt, J.\ \etal 2001, The FORS Deep Field, Reviews in
        Modern Astronomy, Vol. 14, R.E.\ Schielicke (Ed.), Astronomische
        Gesellschaft}
%\bibitem{hellaby06}{Hellaby, C.\ 2006, \mnras, 370, 239}
\bibitem{holanda1}{Holanda, R.F.L., Lima, J.A.S., and Ribeiro, M.B.\ 2010,
        \apjl, 722, L233, arXiv:1005.4458}
\bibitem{holanda2}{Holanda, R.F.L., Lima, J.A.S., and Ribeiro, M.B.\ 2011,
        \aap, 528, L14, arXiv:1003.5906}
\bibitem{holanda3}{Holanda, R.F.L., Lima, J.A.S., and Ribeiro, M.B.\ 2012,
        \aap, 538, A131, arXiv:1104.3753}
\bibitem{iribarrem12}{Iribarrem, A.S., Lopes, A.R., Ribeiro, M.B., and
        Stoeger, W.R.\ 2012, \aap, 539, A112, arXiv:1201.5571, (\textbf{Ir12})}
\bibitem{iribarrem13}{Iribarrem, A.\ \etal 2013, \aap, 558, A15,
        arXiv:1308.2199}
\bibitem{iribarrem14}{Iribarrem, A.\ \etal 2014, \aap, 563, A20,
        arXiv:1401.6572}
\bibitem{CNOC2}{Lin, H.\ \etal 1999, \apj, 518, 533-561}
\bibitem{martinez-saar02} {Mart\'{\i}nez, V.J., and Saar, E.\ 2002, Statistics
        of the Galaxy Distribution (Chapman \& Hall/CRC)}
\bibitem{lopes2014}{Lopes, A.R., and Iribarrem, A., and Ribeiro, M.B., and
	Stoeger, W.R.\ 2014, \aap, 572, A27, arXiv:1409.5409}
\bibitem{p87}{Pietronero, L.\ 1987, Physica A, 144, 257}
\bibitem[2006]{pk06} Pleba\'{n}ski, J., and Krasi\'{n}ski, A.\ 2006,
        An Introduction to General Relativity and Cosmology (Cambridge University Press)
\bibitem{juracy}{Rangel Lemos, L.J., and Ribeiro, M.B.\ 2008,
        \aap, 488, 55-66, arXiv:0805.3336}
\bibitem{rib95}{Ribeiro, M.B.\ 1995, \apj , 441, 477-487,
        arXiv:astro-ph/9910145}
\bibitem{rib01}{Ribeiro, M.B.\ 2001, Gen.\ Rel.\ Grav., 33, 1699-1730,
        arXiv:astro-ph/0104181}
\bibitem{paper2}{Ribeiro, M.B.\ 2005, \aap, 429, 65-74, arXiv:astro-ph/0408316}
\bibitem{paper1}{Ribeiro, M.B., and Stoeger, W.R.\ 2003,
        \apj, 592, 1-16, arXiv:astro-ph/0304094 (\textbf{RS03})}
\bibitem{schechter}{Schechter, P.\ 1976, \apj, 203, 297-306}
\bibitem{sg}{Sparke, L.S., and Gallagher, J.S.\ 2000, Galaxies in
        the Universe (Cambridge University Press)}
\bibitem{sylosetal98} Sylos Labini, F.\ \etal 1998, \physrep, 293, 61-226
\bibitem{tegmark04} Tegmark, M.\ \etal 2004, \apj, 606, 702-740
\bibitem{tegmark06} Tegmark, M.\ \etal 2006, \prd, 74, 123507
\bibitem{sh}{Teles, S., Lopes, A.R., Ribeiro, M.B.\ 2021, Phys.\ Lett.\ B,
813, 136034, arXiv:2012.07164}
\end{thebibliography}
\end{document}